\documentclass[%
showkeys,
showpacs,
reprint,
superscriptaddress,
nofootinbib,
nobibnotes,
 amsmath,amssymb,
aps,
prc,
]{revtex4-2}
\usepackage{amssymb}
\usepackage{graphicx}% Include figure files
\usepackage{dcolumn}% Align table columns on decimal point
\usepackage{bm}% bold math
\usepackage{hyperref}
\usepackage{breakurl}

\begin{document}

\preprint{Version 1}

\title{Stimulated perturbation on the neutron flux distribution in the mutually-dependent source-to-absorber geometry}

\author{Ateia~W.~Mahmoud}\email{atia.mahmoud@gmail.com} \email{ateia.mahmoud@eaea.org.eg}
\address{Physics Department, Faculty of Science, Ain Shams University, Cairo, Egypt.}
\address{Reactor Physics Department, Nuclear Research Center, Egyptian Atomic Energy Authority, Cairo 13759, Egypt.}

\author{Elsayed~K.~Elmaghraby}\email[Corresponding Author: ]{e.m.k.elmaghraby@gmail.com}
\email{elsayed.elmaghraby@eaea.org.eg}
\address{Experimental Nuclear Physics Department, Nuclear Research Center, Egyptian Atomic Energy Authority, Cairo 13759, Egypt.}

\author{A. H. M. Solieman}
\email{ahmedhassanms@gmail.com}
\address{Experimental Nuclear Physics Department, Nuclear Research Center, Egyptian Atomic Energy Authority, Cairo 13759, Egypt.}

\author{E.~Salama}\email{e\_elsayed@sci.asu.edu.eg}
\address{Basic Science Department, Faculty of Engineering, The British University in Egypt (BUE), Cairo, Egypt.}

\author{A.~Elghazaly}\email{an\_4558@yahoo.com}
\address{Reactor Physics Department, Nuclear Research Center, Egyptian Atomic Energy Authority, Cairo 13759, Egypt.}

\author{S.~A.~El-fiki}\email{soadelfiki@sci.asu.edu.eg}
\address{Physics Department, Faculty of Science, Ain Shams University, Cairo, Egypt.}

 % but any date may be explicitly specified

\begin{abstract}
The complexity of the neutron transport phenomenon throws its shadows on every physical system wherever neutron is produced or absorbed. The Monte Carlo N-Particle Transport Code (MCNP) was used to investigate the flux perturbations in the neutron field caused by an absorber. The geometry of the present experiment was designed to reach a simulation of an isotopic neutron field. The neutron source was a ${}^{241}$AmBe with the production physics of neutrons is dependent only on alpha-beryllium interaction and is independent of what happened to the neutron after it was generated. The geometries have been designed to get a volume of uniform neutron densities within a spherical volume of radius 15 cm in every neutron energy group up to 10 MeV. Absorbers of different dimensions were placed within the volume to investigate the field perturbation. Different neutron absorbers were used to correlate the phenomenon to the integral cross section of the absorber. Flux density inside and outside the absorber samples was determined, while the spatial neutron flux distribution produced by the AmBe source without absorber was taken as a reference. This study displayed that absorbers of various dimensions perturb the neutron field in a way that is dependent on the absorption and scattering cross-sections, particularly in the neutron resonance region. Unlike the simple picture of reducing the number density of neutrons, the perturbation was found to influence the moderation of neutrons in the medium, significantly above 1 MeV.
\end{abstract}

\keywords{Dependent source-to-absorber geometry; Neutron flux perturbation; MCNP simulation.}

\pacs{
28.20.Gd %Neutron transport: diffusion and moderation
,
25.40.Dn % Elastic neutron scattering
,
25.40.Ep %Inelastic proton scattering
,
25.40.Fq %Inelastic neutron scattering
,
28.20.-v % Neutron physics
,
28.20.Cz %Neutron scattering
,
{28.41.-i} % {Fission reactors}\sep
,
{28.41.Pa} % {Moderators}\sep
,
{29.25.Dz} % {Neutron sources}.
}

\maketitle

%\tableofcontents

\section{Introduction}

The sources of neutrons are well-known, i.e. reactors, neutron generators, and isotopic neutron sources are the most valuable sources. There are two primary geometries in neutron physics to benefit from neutron sources: beam geometry, in which the source is independent of what happened to the neutron after it has been generated, and field geometry, where the source yield is dependent on what happened to the neutron after it is generated. The reactor's core is field geometry because the core itself uses the produced neutrons to sustain the chain reaction. However, if we take a neutron channel incident to our material, that is no longer field geometry. It is beam geometry because the extracted neutron will be thrown away if it is not absorbed by the sample, for example. Other neutron sources seem to be beam geometries because the physics of neutron production is independent of the neutron itself. However, this may be erroneous if the geometry forces the neutrons to bounce back and forth throughout the source-to-absorber pile.

In a recent set of studies, our group had observed that even the sources we thought to have independence between the neutron flux and the physics at which it is produced can be actually perturbed due to an external or internal perturbation source. In the pile studied by  Tohamy et al. \cite{TohamyElmaghrabyComsan2019162387}, the external moderator scatters the neutrons back and forth to the irradiation chamber so that the overall flux in the system decreases once the neutron is absorbed. Also, in the geometry of work done by Elmaghraby et al. \cite{Elmaghraby2019NPA}, the cooling water of the target moderates the neutrons and reflects them back to the accelerator target. Finally, the geometry of the experiment carried out by Alabyad et al. \cite{AlabyadElaalHassaninElmaghraby2020108947} where the physics of neutron production itself is influenced due to the existence of feedback through ${}^{7}$Li(p,n)${}^{7}$Be which can absorb a thermal neutron and form ${}^{8}$Be which decays to two helium atoms. All these remarks, which caused difficulties and interferences in the addresses measurements, can be explained as a stimulated perturbation on the neutron flux distribution in the mutually-dependent source-to-absorber pile.

\noindent  The sensitivity expressions derived from perturbation formulas using the transport equation in Beckurts and Wirtz \cite{beckurtsWirtz1964} are easier to use than their integral counterparts in MCNP; however, they are limited to Maxwellian neutron distributions. Among the parameters that determine such a phenomenon is the neutron flux perturbation factor defined as the ratio between the neutron fluxes present in the field whenever a material exists to the unperturbed neutron flux that cannot be measured without actual perturbation. One of the primary problems that have been encountered in the experimental measurement of flux perturbations is the determination of the unperturbed flux \cite{walker1963thermal,TohamyElmaghrabyComsan2019162387,Elmaghraby2018PhysScr,TohamyElmaghrabyComsan2021045304}. During the last decade, a great deal of research effort (both theoretical and experimental) has been directed towards a better understanding of the spectral characteristics of in-core neutron-noise measurements in BWRs \cite{analytis1982analysis}. The main goal of such neutron noise analysis is to reach decisions concerning the behavior of the noise sources by studying the neutron \cite{behringer1979linear}. These research efforts (both theoretical and experimental) were initially reported by Seifritz \cite{seifritz1972analysis} and \cite{seifritz1973load}, studies becomes extensive in later work in Refs. \cite{analytis1982analysis,Laggiard1995124,Antonopoulos-Domis1999337,Yamamoto2021190}. The thermal neutron flux perturbation factor is a significant source of error in the case of the absolute thermal neutron activation technique \cite{buczko1978simple}. Several methods use the thermal neutron activation analysis compared to fast neutron activation related to the high-activation cross-sections and the absence of interfering reactions that exist in case of high energy \cite{csikai1987crc}. The severe limitation of these techniques is the perturbation of thermal neutron flux by strongly absorbing and extended samples \cite{csikai2002studies} and, as we shall emphasize herein, the possible perturbation of fast neutron itself.

In the present work, we emphasized the role of neutron field perturbations using hypothetical simulations and the Monte Carlo N-Particle Transport Code (MCNP). Experimental determination of such perturbations is a difficult task. The dependence of neutron field perturbations on the energy domain of the neutrons shall be investigated for different materials of different dimensions to reach a conclusion about the perturbation of the neutron field in a manner that depends on the absorption and the scattering cross-sections of these materials.

\section{Materials and Methods}
\subsection{Mesh and Tallies}

\begin{figure*}[htbp]
  \centering
  \includegraphics[width=0.99\linewidth]{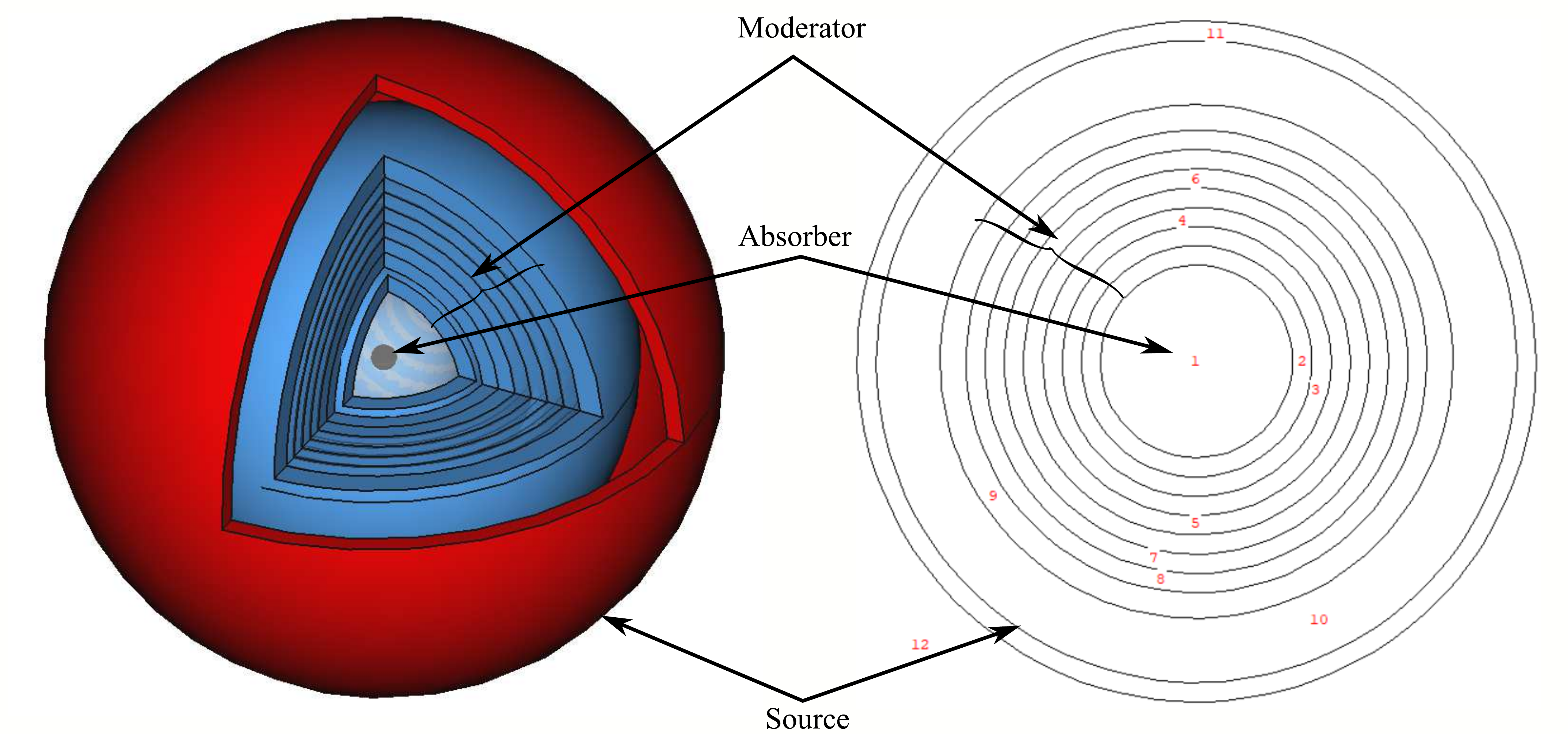}
  \caption{A schematic diagram of the designed simulation using MCNP. left panel is a 3D illustration while right panel is a 2D cross-section with volume number.}\label{Fig:Geometry}
\end{figure*}

\noindent MCNP is a general-purpose Monte Carlo N--Particle code usually used for neutron transport, that treats the three-dimensional configuration of sources and materials in geometric cells \cite{Thompson1979Mcnp}, and expresses the transport of the particle in many parameters including count and energy. In our model, the neutron flux distributions inside and outside the samples from the isotropic ${}^{241}$Am-Be source were calculated using MCNP version 5 (ENDF-VI) and its associated libraries. Our indicator for flux perturbation is the number of energy-dependent tracks within every spherical shell of variable radii ($r$) centered at the center of mass of the pile shown in Fig. \ref{Fig:Geometry}.  The number of tracks within specific spheres of radius r was obtained by dividing the energy of neutrons into 22 domains, i.e. 0-0.01 eV, 0.01-0.025 eV, 0.025-0.05 eV, 0.05-0.1 eV, 0.1-1 eV, 1-10 eV, 10 eV-100 eV, 100 eV-1 keV, 1-10 keV, 10-100 keV, 100-500 keV, 500-800 keV, 800 keV-1 MeV, 1-2 MeV, 2-3 MeV, 3-4 MeV, 4-5 MeV, 5-6 MeV, 6-7 MeV, 7-8 MeV, 8-9 MeV, and 9-10 MeV. The track length estimate of the neutron flux, averaged over mesh cells in units of neutron/cm${}^{2}$, was calculated using the FMESH card.

\subsection{Statistical variance reduction }

The variance reduction technique, the population control method, was used to achieve the lowest relative errors and pass the statistical tests. For employing this technique, the moderating volume was split into several sections based on the importance of neutrons. The neutrons outside the pile have been killed. With 10${}^{8}$ histories, the applicable variance reduction technique created relative errors in the output results less than 2\%. To reach this value, we have redesigned the MCNP geometry: a halo sphere source (No. 11) made of an Am-Be were placed around a set of successive spherical shells spheres of moderating medium (Nos. 2-9), as in Fig. \ref{Fig:Geometry}. Each moderating medium between its outer and inner boundaries was filled with water as a moderator without any volume between the preceding and subsequent moderating medium. The halo sphere source grantee equals currents of neutrons from all coordinates directions. At the same time, the moderating medium shall induce homogenization of the neutrons by multiple scattering to cover the 4$\pi $ dimensions around its internal void (volume no. 1).

\section{Results and discussion}

The geometry of pile in MCNP was challenging to build using existing experimental geometries (cf. Refs. \cite{TohamyElmaghrabyComsan2019162387,Elmaghraby2019NPA,AlabyadElaalHassaninElmaghraby2020108947}) because of two contradicting reasons:
\begin{itemize}
\item the neutrons need to form a uniform neutron field within the sample position,
\item the field needs to be isotropic.
\end{itemize}
\noindent The uniform field means to have a constant fluence rate (flux) at every position in the volume at which the absorber exists. Isotropic field requires neutrons to be incident from all directions with the same number density.   To overcome this issue, we designed the MCNP geometry as follows. The spherical shell of the Am-Be source with 3 cm in thickness and 50 cm in radius (inner) is placed around a halo sphere of moderating medium. Neutron Emission Probabilities for ${}^{241}$AmBe neutron source (W'(En)) in units lethargy (ul${}^{-1}$) can be found in Ref. \cite{TohamyElmaghrabyComsan2019162387}. The moderating medium shall induce homogenization of the neutrons by multiple scattering to cover the 4$\pi $ direction around its internal void. The halo sphere has a radius of the outer part of 40 cm. The radius of the inner part is 15 cm, leaving a thickness of 25 cm of moderating medium that is enough to moderate the standard emission spectrum of Am-Be source lower energies. This geometry of the source makes the center of mass of the pile coincide with the center of mass of the neutron number density. The full spectrum doesn't have to reach the Maxwellian distribution (with T=300 K), but rather it is beneficiary for the present study to reach partially moderated neutron spectrum. The moderating medium between the outer and inner boundaries of the halo spheres is assumed to be water. The samples were placed at the center of mass of the system within the inner part of the void. The absorbers are assumed to be spheres of one of the illustrated absorbers with different radii from 1 to 6 cm, each absorber was placed inside that inner part, as shown in Fig. \ref{Fig:Geometry}.

\begin{figure}[htbp]
  \centering
  \includegraphics[width=\linewidth]{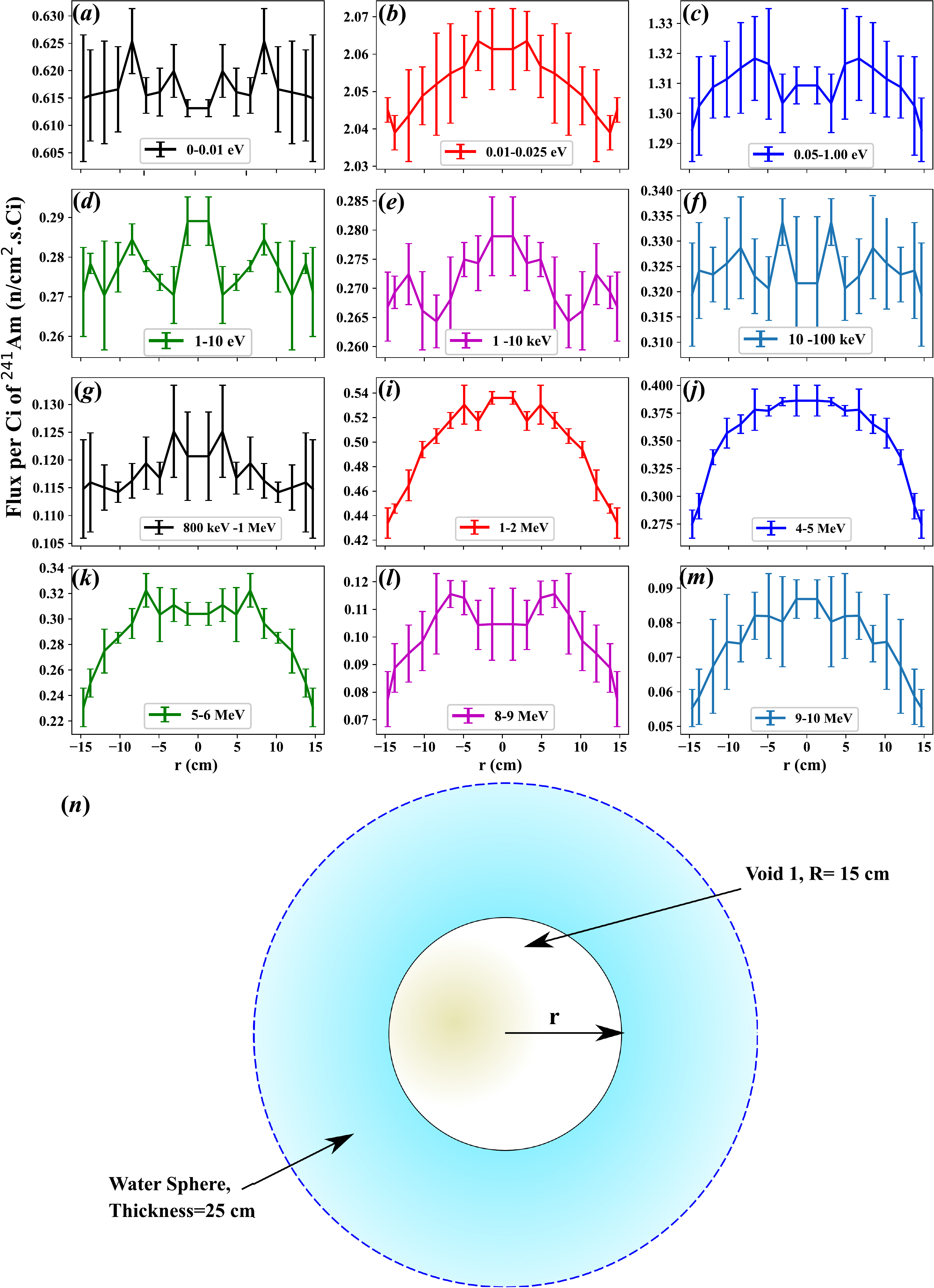}
  \caption{The special distribution of neutrons in 12 energy domains along the 22 energy domains as a function of the radial distance within the void 1. Other domains were disregarded to enable clear illustrations.}\label{Fig:SpacialEnergyDistribution}
\end{figure}

\begin{figure}[htbp]
  \centering
  \includegraphics[width=0.7\linewidth]{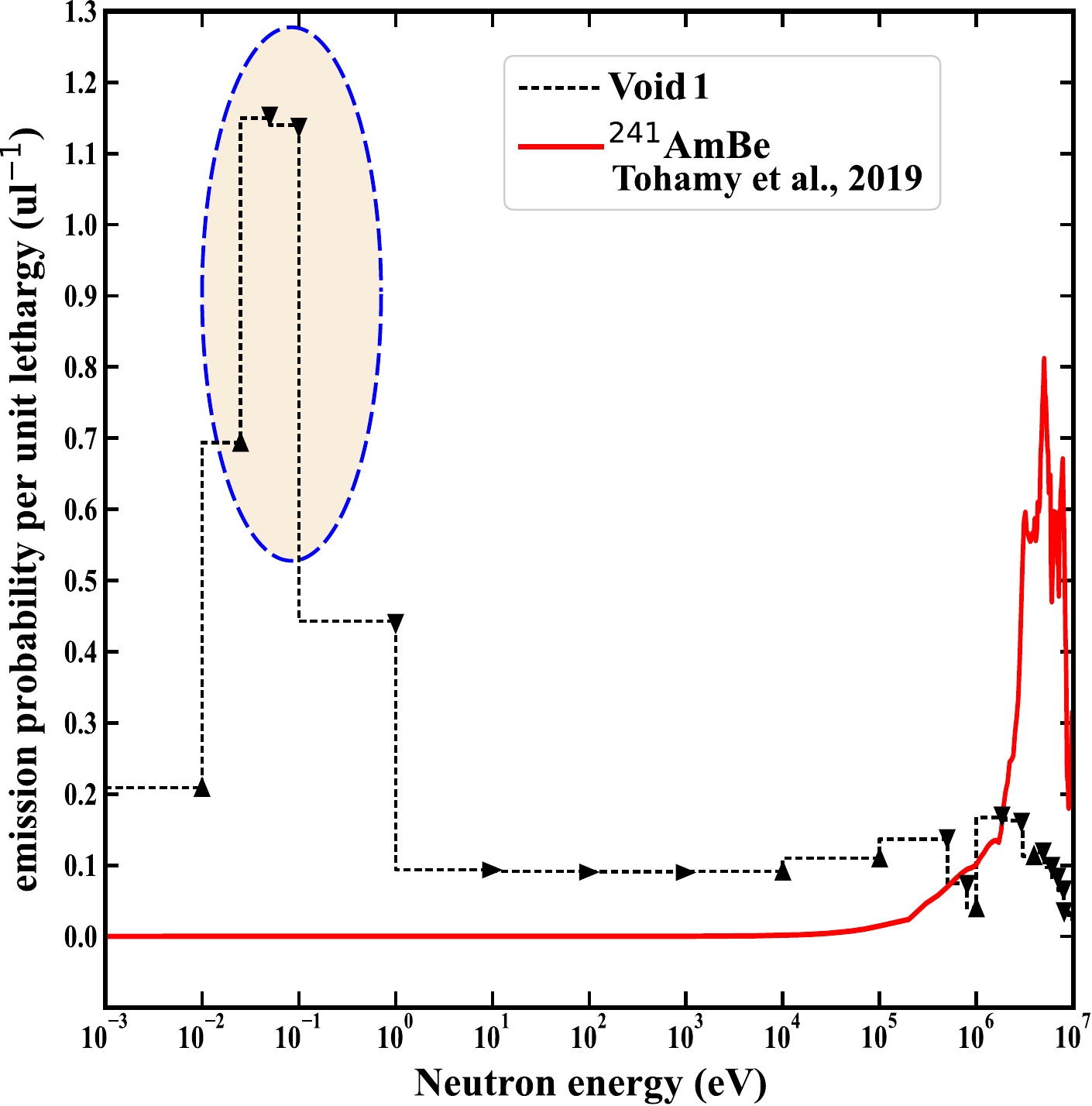}
  \caption{The average neutron energy distribution in the center of the void 1. The unit is the probability of emission divided by the unit lethargy of the preceding energy domain. The distribution is compared to the standard emission probability of neutrons from {${}^{241}$}AmBe neutron source taken from Ref. \cite{TohamyElmaghrabyComsan2019162387}. The probability per unit lethargy exceeds 1 in such case in which the difference in ln(energy) is small enough.}\label{Fig:averageenergydistribution}
\end{figure}

Following the geometry shown in Fig. \ref{Fig:Geometry}, the spatial distribution of neutrons in the void (sphere no. 1) is calculated at different distances from the pile's center of mass to ratify the uniformity of neutron number density throughout volume no. 1. Figure \ref{Fig:SpacialEnergyDistribution} illustrates this distribution for the 12 out of 22 simulated energy domains. Typically, after moderation, most of the neutrons have less than 1 eV of energy. Nearly all spatial distributions have uniform neutron number densities overall the volume of the void, especially the first seven domains from low energy up to 1 MeV.
Higher energy neutrons have a slightly higher density around the center of the void. This profile is a consequence of the spherical symmetry of the pile in which neutron number density around the center-of-curvature of the pile as whole represents the region for increasing number density of neutrons (either un-moderated or partially moderated neutrons coming from the surrounding source). The average energy distribution is illustrated in Fig. \ref{Fig:averageenergydistribution}.

Separately, the calculations were repeated with spherical absorbers of different diameters from 2 cm to 12 cm are located at the center of the void no. 1. These absorbers are made of either strong absorber (indium, gold, or mercury) or weak absorbers (iron or zinc), and selected to have different thermal neutrons cross sections and resonance integrals \cite{TohamyElmaghrabyComsan2020109340,Elmaghraby2019PhysScrCode,Elmaghraby2016Shape}.

\subsection{Strong absorbers}
\begin{figure}[htbp]
  \centering
  \includegraphics[width=0.99\linewidth]{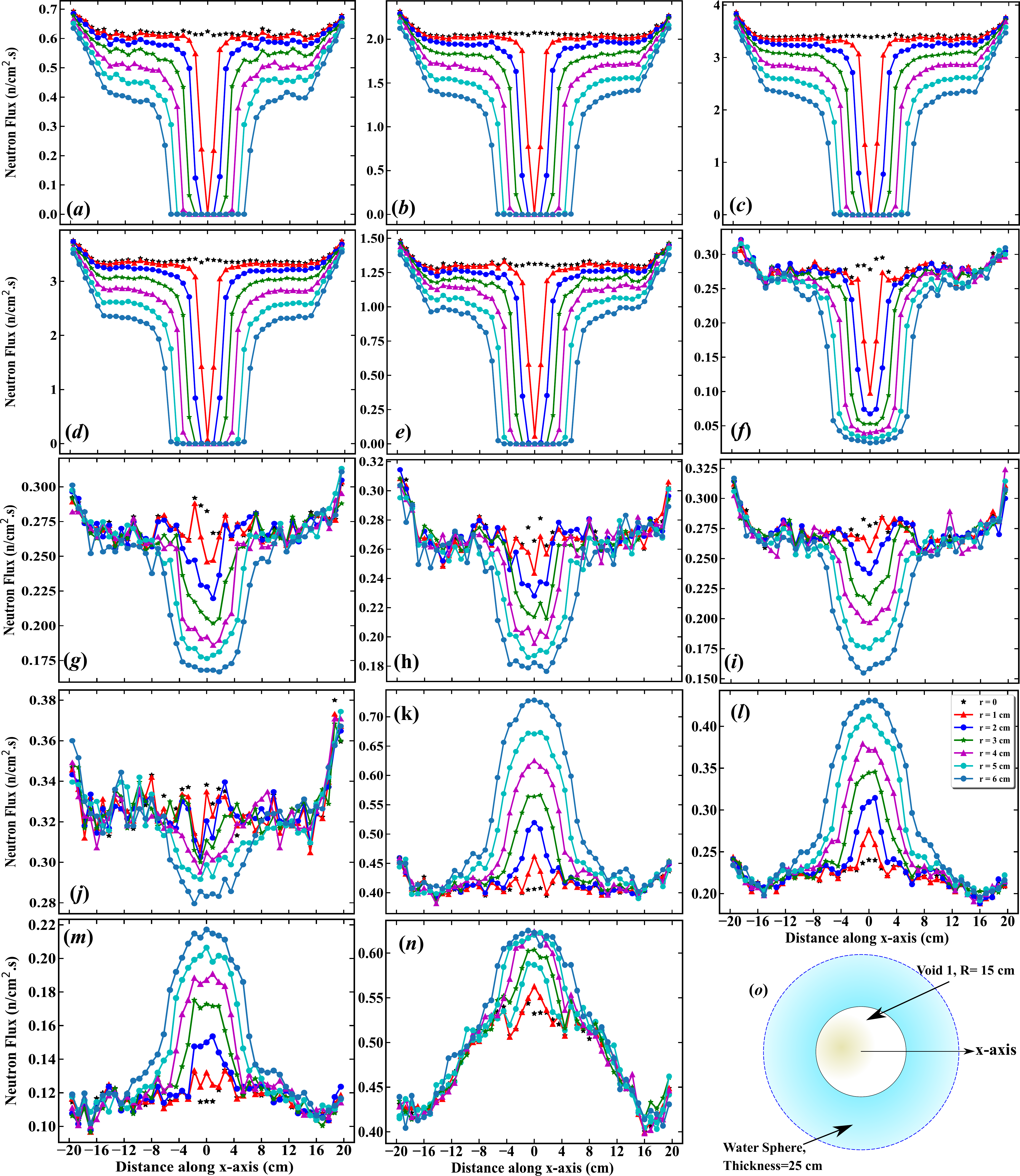}
  \caption{Neutron flux of in the vicinity of indium absorbers having different dimensions at 14 different domains of energies. (a) 0-0.01 eV, (b) 0.01-0.025 eV, (c) 0.025-0.05 eV, (d) 0.05-0.1 eV, (e) 0.1-1 eV, (f) 1-10 eV, (g) 10 eV-100 eV, (h) 100eV-1 keV, (i) 1-10 keV, (j) 10-100 keV, (k) 100-500 keV, (l) 500-800 keV, (m) 800 keV-1 MeV, and (n) 1-2 MeV. (o) The direction of the x axis coordinate relative to the setup as illustrated in Fig. \ref{Fig:Geometry}}\label{Fig:Indium1}
\end{figure}

Indium has two isotopes $^{115} $In (IA = 0.95719) and $^{113} $In (IA = 0.04281) \cite{Meija2016293IA}. For $^{115} $In, the thermal neutron absorption cross section is $\sigma _{0} =$202.22 b while the resonance integral in pure 1/E-neutron spectrum is $I_{0} =$3211 b \cite{Sukhoruchkin2009book,Mughabghab2006Book,Elmaghraby2019PhysScrCode,Elmaghraby2016Shape}. However, the thermal neutron cross section and resonance integral for the specific reaction $^{115} $In(n,$\gamma $)$^{116m} $In are 162.6 b and 2585 b \cite{Elmaghraby2018PhysScr}, respectively. This value makes elemental indium a strong absorber for neutrons. $^{113} $In, however, has small value of isotopic abundance and much smaller thermal neutron cross section and resonance integral, i.e. $\sigma _{0} =$12.13  b and $I_{0} =$332.2 b \cite{Sukhoruchkin2009book,Mughabghab2006Book,Elmaghraby2019PhysScrCode,Elmaghraby2016Shape}, so $^{113} $In has a limited contribution to the overall macroscopic cross section of the element. In our MCNP simulation,  positioning indium absorber in the center of mass of the pile causes a reduction in the total thermal flux within the void no. 1; see Fig. \ref{Fig:Indium1}a,b. As the size of the absorber increases, the macroscopic cross section increases, and consequently the reductions become more pronounced as obviously shown in Fig. \ref{Fig:Indium1}a,b. Noticeably, this behavior of reduction in the total flux persists within energy epithermal domains, as illustrated in Fig. \ref{Fig:Indium1}c (0.025-0.05 eV), Fig. \ref{Fig:Indium1}d (0.05-0.1 eV), Fig. \ref{Fig:Indium1}e (0.1-1 eV), Fig. \ref{Fig:Indium1}f (1-10 eV), Fig. \ref{Fig:Indium1}g (10 eV-100 eV), Fig. \ref{Fig:Indium1}h (100 eV-1 keV), Fig. \ref{Fig:Indium1}i (1-10 keV), and Fig. \ref{Fig:Indium1}j (10-100 keV), with decrease in the strength of the perturbation as the energy of neutron reaches the 100 keV.

The sign of the flux perturbation is inverted at energies greater than 100 keV, i.e. the effect of perturbation appeared inverted at high energy from domains as shown in Fig. \ref{Fig:Indium1}k (100-500 keV), Fig. \ref{Fig:Indium1}l (500-800 keV), Fig. \ref{Fig:Indium1}m (800 keV-1 MeV), and Fig. \ref{Fig:Indium1}n (1-2 MeV), where the neutron flux is enhanced in these energy ranges. {The word enhanced here is misleading because what happens actually is that these extra neutrons represent neutrons that were partially moderated or not moderated at all in the field as a result of sample perturbation and allowed to go deeper in the absorber due to the small absorption cross section of these neutron within the surface layers of the absorber}. The scattering within the absorber increases the fraction of neutrons in this range (100 keV to 2 MeV) at the expense of the number of neutrons in higher energy domains.

\begin{figure}[htbp]
  \centering
  \includegraphics[width=0.99\linewidth]{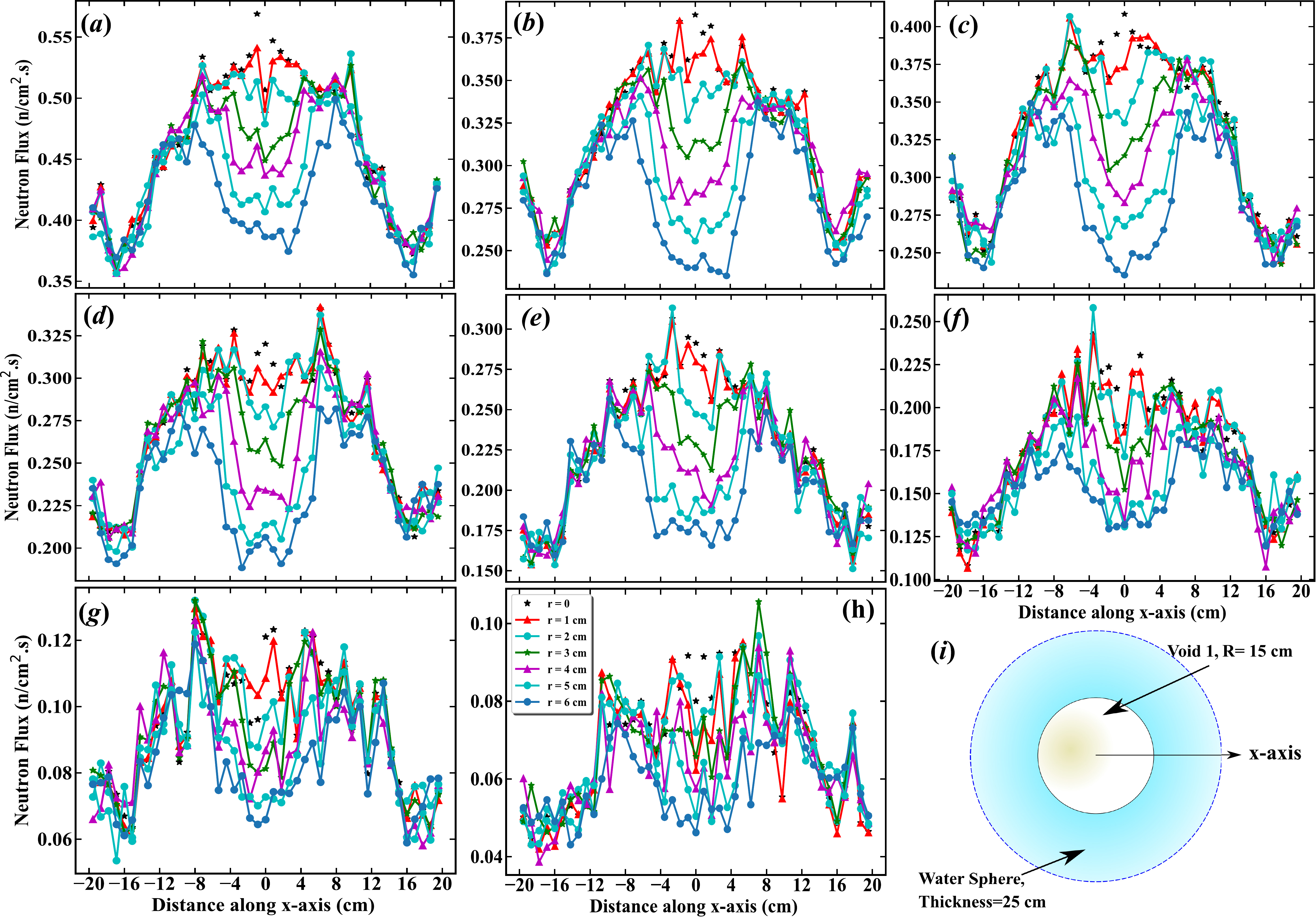}
  \caption{Neutron flux of in the vicinity of indium absorbers having different dimensions at 8 different domains of energies. (a) 2-3 MeV, (b) 3-4 MeV, (c) 4-5 MeV, (d) 5-6 MeV, (e) 6-7 MeV, (f) 7-8 MeV, (g) 8-9 MeV, and (h) 9-10 MeV. Part (i) as in Fig. \ref{Fig:Indium1}o.}\label{Fig:Indium2}
\end{figure}

At energies greater than 2 MeV, i.e. the energy domains 2-3 MeV (Fig. \ref{Fig:Indium2}a), 3-4 MeV (Fig. \ref{Fig:Indium2}b), 4-5 MeV (Fig. \ref{Fig:Indium2}c), 5-6 MeV (Fig. \ref{Fig:Indium2}d), 6-7 MeV (Fig. \ref{Fig:Indium2}e), 7-8 MeV (Fig. \ref{Fig:Indium2}f), 8-9 MeV (Fig. \ref{Fig:Indium2}g), and 9-10 MeV (Fig. \ref{Fig:Indium2}g), another phenomenon acted. There are a kind of reduction in neutron flux which becomes more pronounced as the size of the indium absorber increases. Of course this happens together with the effect of the non-uniformity of spatial neutron distribution, as illustrated in Fig. \ref{Fig:SpacialEnergyDistribution}. The associated context is the absorption of these fast neutrons at the center of the absorber in terms of another regime of interaction specifically the inelastic scattering with indium isotopes, e.g. $^{115} $In(n,n')$^{115m,g} $In, which is more likely to occur in this energy range \cite{TohamyElmaghrabyComsan2020109340}. Hence, the perturbation becomes affected by fast neutron absorption within the absorbing sphere and re-emission of inelastically scattered neutrons with lower energies..

\begin{figure}[htbp]
  \centering
  \includegraphics[width=0.99\linewidth]{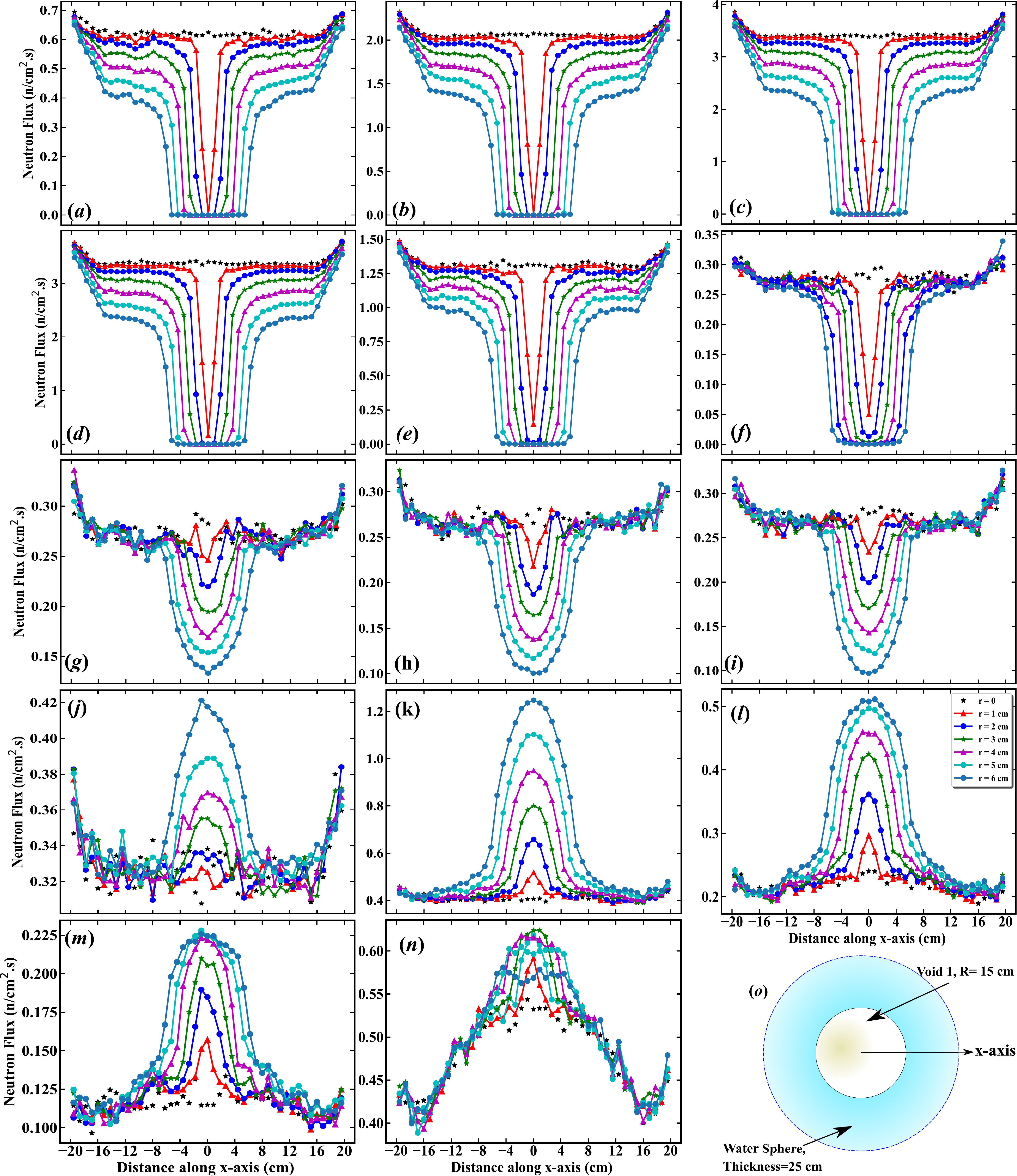}
  \caption{Neutron flux of in the vicinity of gold absorbers having different dimensions at 14 different domains of energies. Constructed as in Fig. \ref{Fig:Indium1}.}\label{Fig:Gold1}
\end{figure}

\begin{figure}[htbp]
  \centering
  \includegraphics[width=0.99\linewidth]{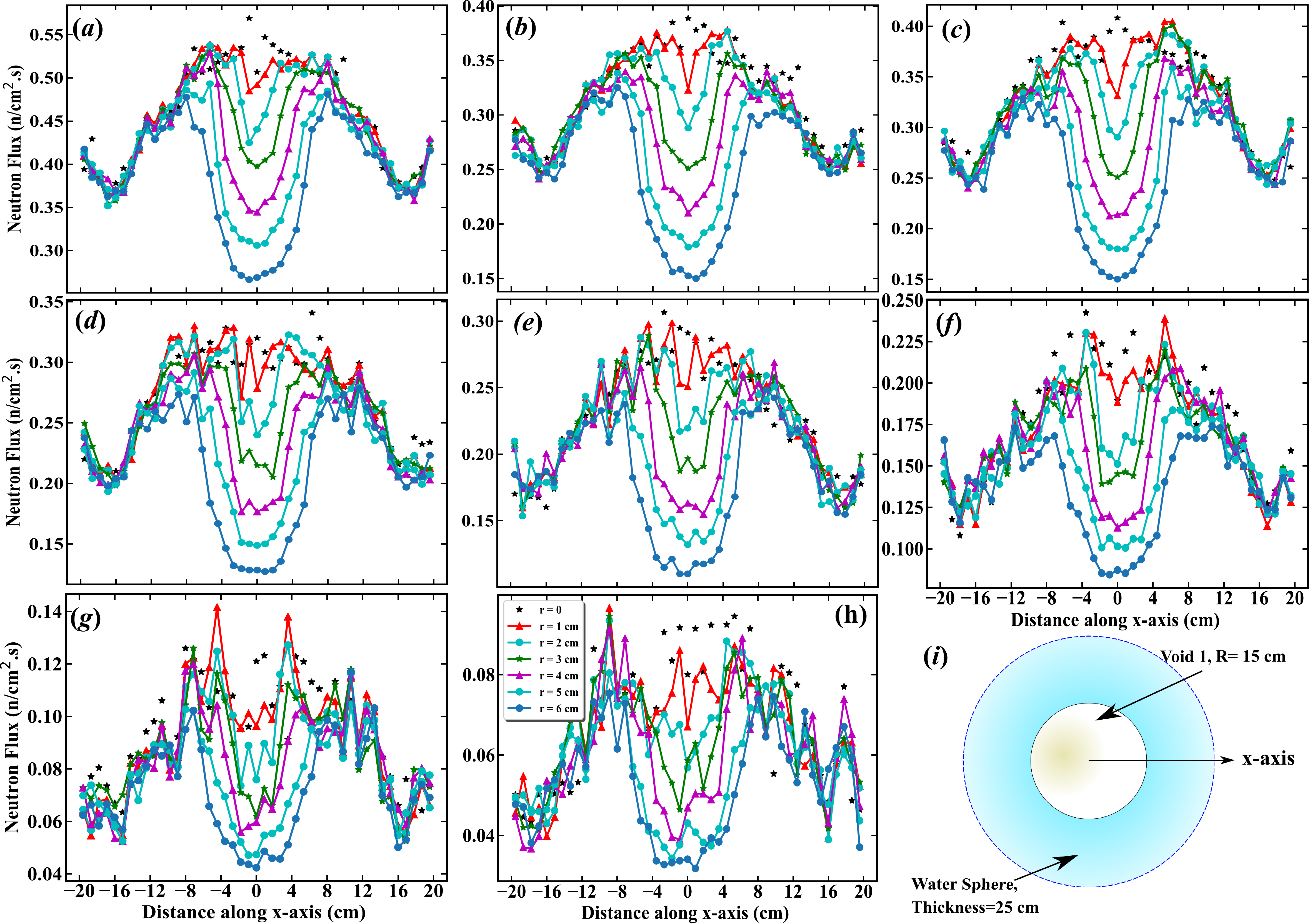}
  \caption{Neutron flux of in the vicinity of gold absorbers having different dimensions at 8 different domains of energies. Constructed as in Fig. \ref{Fig:Indium2}.}\label{Fig:Gold2}
\end{figure}

Gold is a mono-isotope element having only $^{197} $Au whose thermal neutron absorption cross section is $\sigma _{0} =$98.65 b while the resonance integral in pure 1/E-neutron spectrum is $I_{0} =$1567 b \cite{Sukhoruchkin2009book,Mughabghab2006Book,Elmaghraby2019PhysScrCode,Elmaghraby2016Shape}. $^{197} $Au, however, has an intense neutron resonance at 4.89 eV with capture width of $\Gamma _{\gamma } $=0.124 eV \cite{Sukhoruchkin2009book,Elmaghraby2019PhysScrCode}. Such large resonance had an implication on the effect of gold on neutron field perturbation as shown in Fig. \ref{Fig:Gold1}a,b. With standing gold absorber in the center of mass of the pile, the same reduction in the total thermal flux within the void 1, as in case of indium, appears.
This behavior of reduction in the total flux persists within energy of epithermal domains, i.e. Fig. \ref{Fig:Gold1}c (0.025-0.05 eV), Fig. \ref{Fig:Gold1}d (0.05-0.1 eV), Fig. \ref{Fig:Gold1}e (0.1-1 eV), Fig. \ref{Fig:Gold1}f (1-10 eV), Fig. \ref{Fig:Gold1}g (10 eV-100 eV), Fig. \ref{Fig:Gold1}h (100eV-1 keV), and Fig. \ref{Fig:Gold1}i (1-10 keV), with decrease in the strength of the perturbation as the energy of neutron reaches the 10 keV. However, the sign of the flux perturbation is inverted at energies greater than 10 keV, unlike the case of In absorber, i.e. the effect of perturbation appeared inverted at high energy from domains 10-100 keV (Fig. \ref{Fig:Gold1}i), 100-500 keV (Fig. \ref{Fig:Gold1}j), 500-800 keV (Fig. \ref{Fig:Gold1}k), 800 keV-1 MeV (Fig. \ref{Fig:Gold1}l), 1-2 MeV (Fig. \ref{Fig:Gold1}m), where the neutron flux is increased in these energy ranges. At these domains of energy, the scattering within the absorber increases the fraction of neutrons in this range (10 keV to 2 MeV) at the expense of the number of neutrons in higher energy domains. At energies greater than 2 MeV, i.e. the energy domains 2-3 MeV (Fig. \ref{Fig:Gold2}a), 3-4 MeV (Fig. \ref{Fig:Gold2}b), 4-5 MeV (Fig. \ref{Fig:Gold2}c), 5-6 MeV (Fig. \ref{Fig:Gold2}d), 6-7 MeV (Fig. \ref{Fig:Gold2}e), 7-8 MeV (Fig. \ref{Fig:Gold2}f), 8-9 MeV (Fig. \ref{Fig:Gold2}g), and 9-10 MeV (Fig. \ref{Fig:Gold2}h), the reduction in neutron flux which becomes more pronounced and increases further as the size of the gold absorber increases due to inelastic scattering of gold.

\begin{figure}[htbp]
  \centering
  \includegraphics[width=0.99\linewidth]{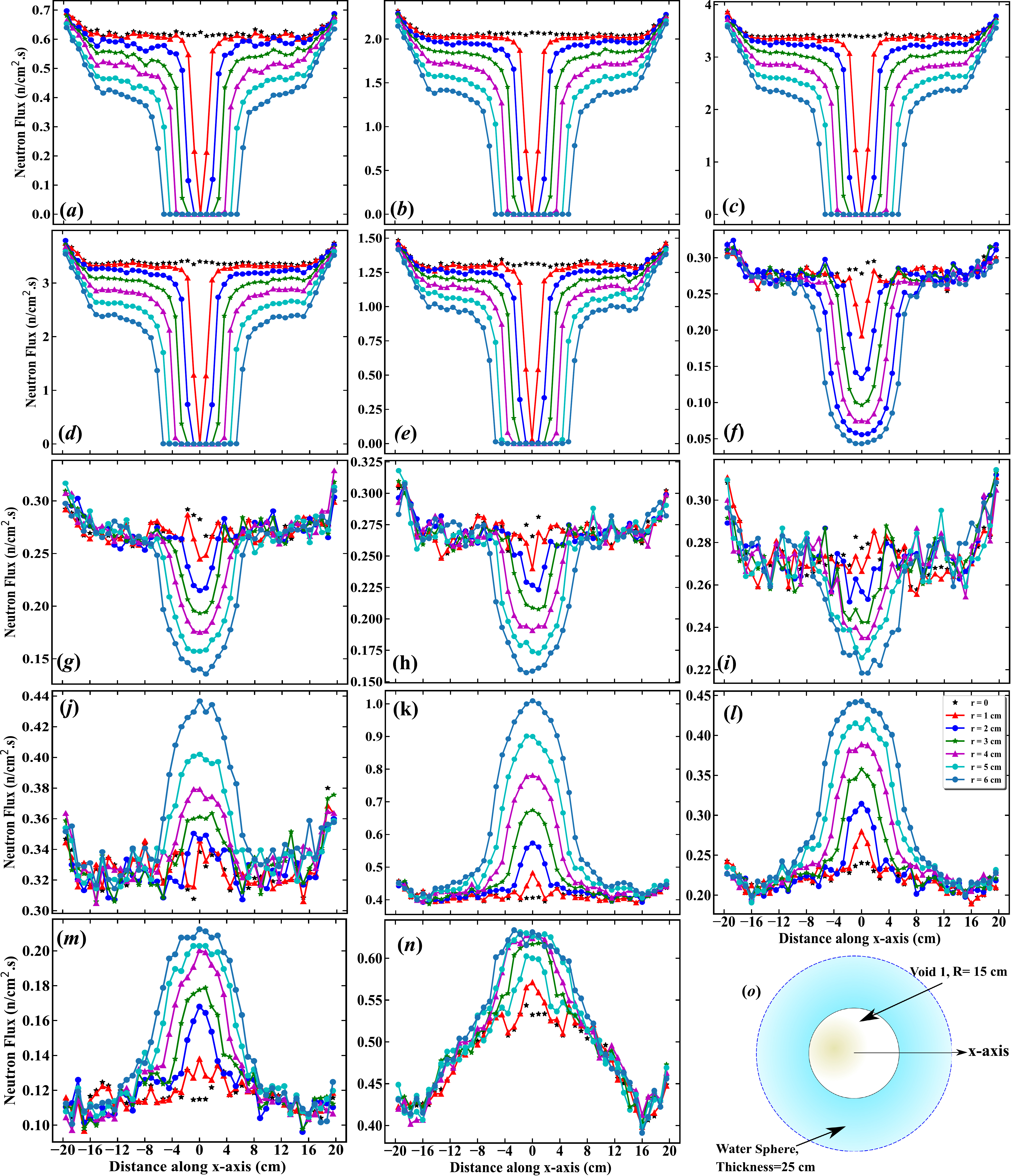}
  \caption{Neutron flux of in the vicinity of mercury absorbers having different dimensions at 14 different domains of energies. Constructed as in Fig. \ref{Fig:Indium1}.}\label{Fig:Mercury1}
\end{figure}

\begin{figure}[htbp]
  \centering
  \includegraphics[width=0.99\linewidth]{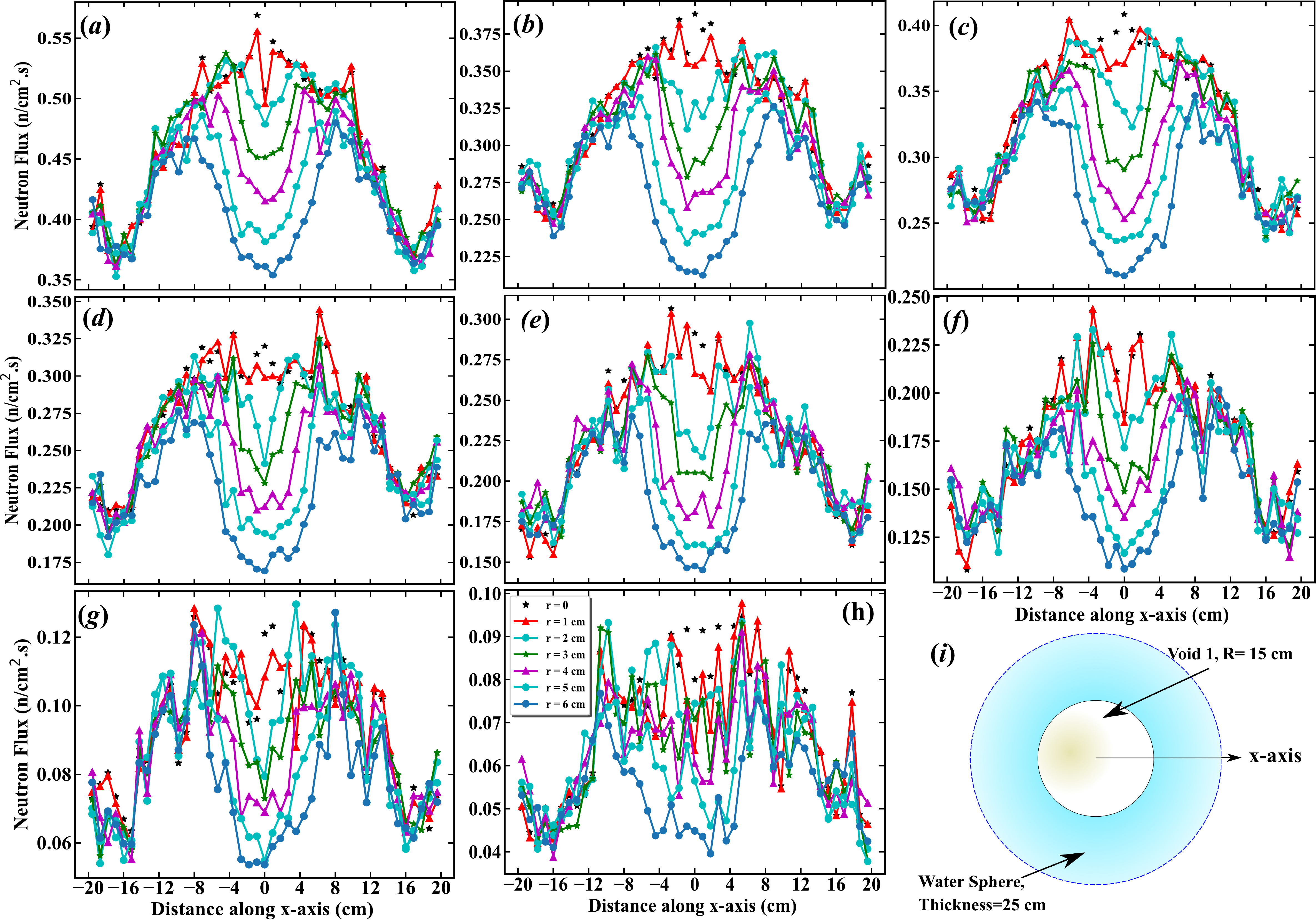}
  \caption{Neutron flux of in the vicinity of mercury absorbers having different dimensions at 8 different domains of energies. Constructed as in Fig. \ref{Fig:Indium2}.}\label{Fig:Mercury2}
\end{figure}

The next strong absorber is mercury. Mercury has 7 stable isotopes, including $^{198} $Hg (IA=0.10038, $\sigma _{o} =$1.98 b, $I_{\gamma } =$74.4 b), $^{199} $Hg (IA=0.16938, $\sigma _{o} =$2149 b, $I_{\gamma } =$408 b), $^{200} $Hg (IA=0.2314, $\sigma _{o} =$1.44 b, $I_{\gamma } =$2.53 b), $^{201} $Hg (IA=0.1317, $\sigma _{o} =$4.90 b, $I_{\gamma } =$33.2 b), $^{202} $Hg (IA=0.29743, $\sigma _{o} =$4.95 b, $I_{\gamma } =$3.07 b), and $^{204} $Hg (IA=0.06818), data are from Refs.  \cite{Meija2016293IA,Sukhoruchkin2009book,Mughabghab2006Book,Elmaghraby2019PhysScrCode,Elmaghraby2016Shape}. The strongest absorber among them is $^{199} $Hg. With positioning mercury absorber in the center of mass of the pile, the same reduction in the total thermal flux within the void 1 as in case of gold; see Fig. \ref{Fig:Mercury1}a,b.
Again, the behavior is a reduction in the total flux within thermal energy and epithermal domains, i.e. 0.025-0.05 eV (Fig. \ref{Fig:Mercury1}c), 0.05-0.1 eV (Fig. \ref{Fig:Mercury1}d), 0.1-1 eV (Fig. \ref{Fig:Mercury1}e), 1-10 eV (Fig. \ref{Fig:Mercury1}f), 10 eV-100 eV (Fig. \ref{Fig:Mercury1}g), 100eV-1 keV (Fig. \ref{Fig:Mercury1}h), and 1-10 keV (Fig. \ref{Fig:Mercury1}i), with decrease in the strength of the perturbation as the energy of neutron reaches the 10 keV. and flux perturbation is inverted at energies greater than 10 keV, i.e. within the domains 10-100 keV  (Fig. \ref{Fig:Mercury1}j), 100-500 keV (Fig. \ref{Fig:Mercury1}k), 500-800 keV (Fig. \ref{Fig:Mercury1}l), 800 keV-1 MeV  (Fig. \ref{Fig:Mercury1}m), and 1-2 MeV (Fig. \ref{Fig:Mercury1}n), where the neutron flux is increased in these energy ranges and in the energy domains 2-3 MeV (Fig. \ref{Fig:Mercury2}a), 3-4 MeV (Fig. \ref{Fig:Mercury2}b), 4-5 MeV (Fig. \ref{Fig:Mercury2}c), 5-6 MeV (Fig. \ref{Fig:Mercury2}d), 6-7 MeV (Fig. \ref{Fig:Mercury2}e), 7-8 MeV (Fig. \ref{Fig:Mercury2}f), 8-9 MeV(Fig. \ref{Fig:Mercury2}g), and 9-10 MeV (Fig. \ref{Fig:Mercury2}h), the of reduction in neutron flux which becomes more pronounced and increases further as the size of the indium absorber increases due to higher energy inelastic with mercury.

\subsection{Weak absorbers}
\begin{figure}[htbp]
  \centering
  \includegraphics[width=0.99\linewidth]{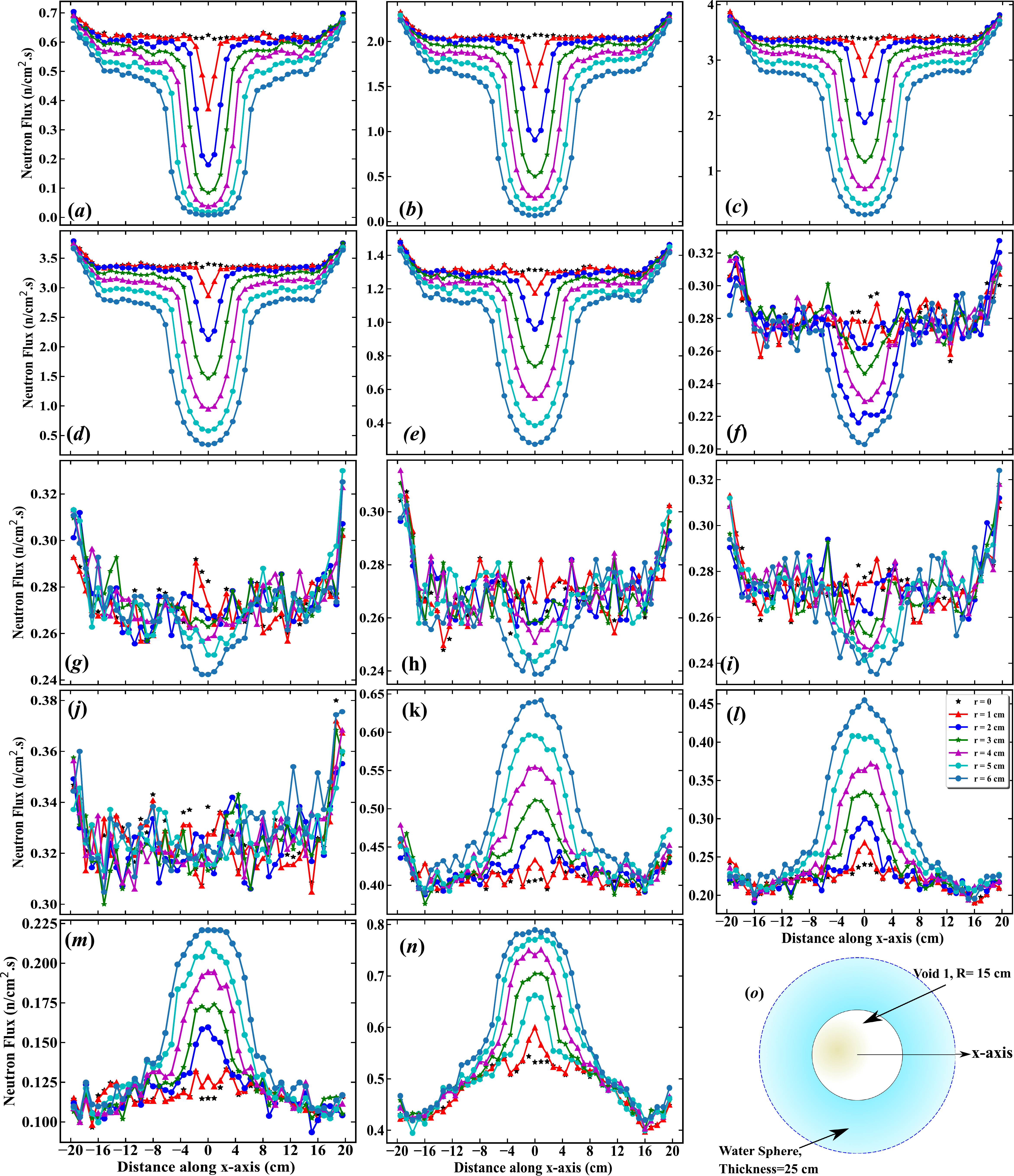}
  \caption{Neutron flux of in the vicinity of iron absorbers having different dimensions at 14 different domains of energies. Constructed as in Fig. \ref{Fig:Indium1}.}\label{Fig:Iron1}
\end{figure}

\begin{figure}[htbp]
  \centering
  \includegraphics[width=0.99\linewidth]{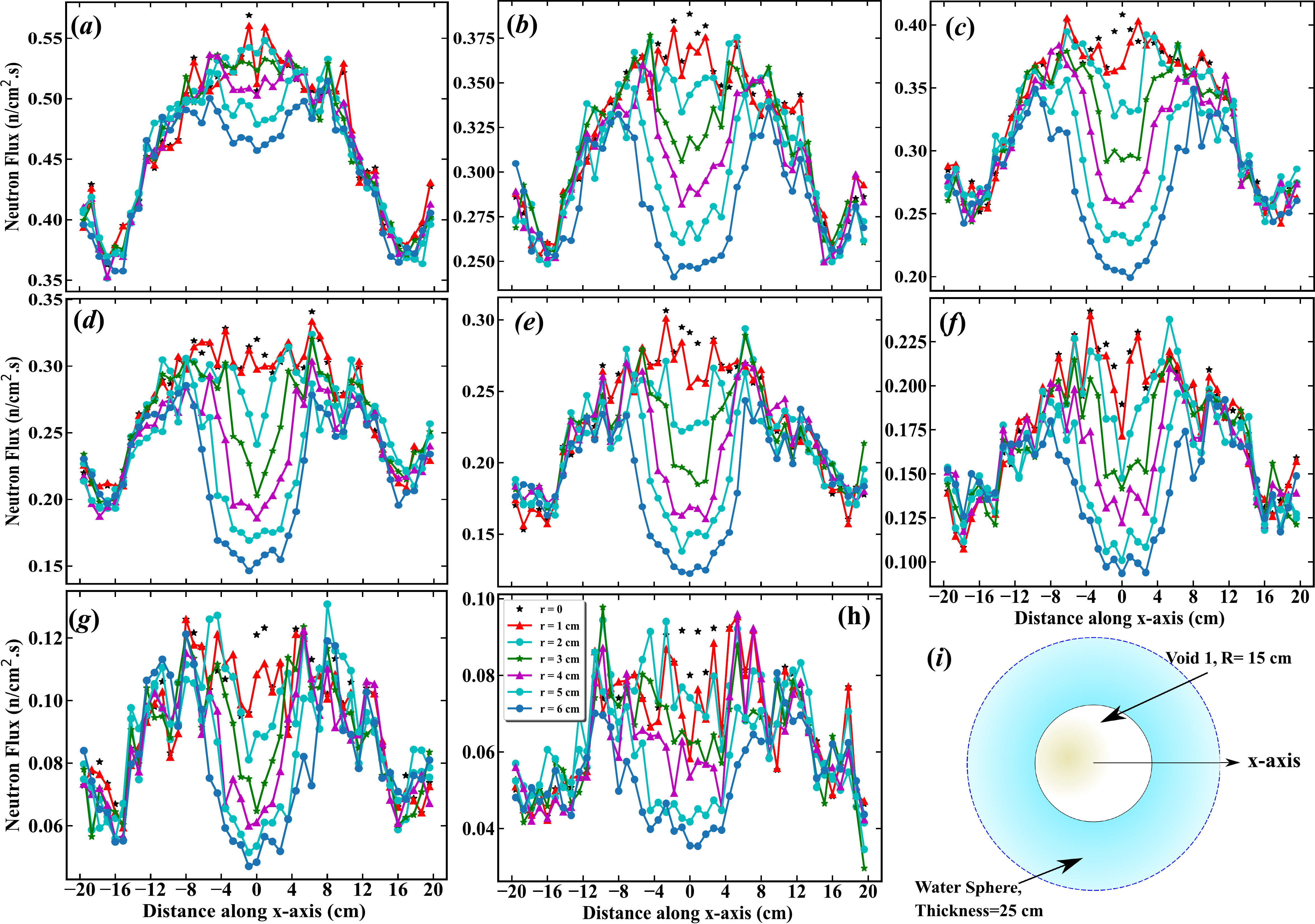}
  \caption{Neutron flux of in the vicinity of iron absorbers having different dimensions at 8 different domains of energies. Constructed as in Fig. \ref{Fig:Indium2}.}\label{Fig:Iron2}
\end{figure}

The second studied case is the weak absorbers such as iron. Iron has four isotopes ($^{54} $Fe (IA=0.05845),  $^{56} $Fe (IA=0.91754), $^{57} $Fe (IA=0.02119) and $^{58} $Fe (IA=0.00282) \cite{Meija2016293IA}. For $^{56} $Fe, the thermal neutron absorption cross section is $\sigma _{0} =$2.59 b while the resonance integral in pure 1/E-neutron spectrum is $I_{0} =$1.27 b \cite{Sukhoruchkin2009book,Mughabghab2006Book,Elmaghraby2019PhysScrCode,Elmaghraby2016Shape}. It is a weak absorber with strong resonance absorptions between 100 keV and 1 MeV; actually begins from 10 keV but with less level densities. This information is reflected back to the neutron field perturbation induced by iron as shown in Figs. \ref{Fig:Iron1}a,b.
The behavior of reduction in the total flux persists within energy domains 0-0.01 eV (Fig. \ref{Fig:Iron1}a), 0.01-0.025 eV (Fig. \ref{Fig:Iron1}b), 0.025-0.05 eV (Fig. \ref{Fig:Iron1}c), 0.05-0.1 eV (Fig. \ref{Fig:Iron1}d), 0.1-1 eV (Fig. \ref{Fig:Iron1}e), 1-10 eV (Fig. \ref{Fig:Iron1}f), 10 eV-100 eV (Fig. \ref{Fig:Iron1}g), 100eV-1 keV (Fig. \ref{Fig:Iron1}h), and 1-10 keV (Fig. \ref{Fig:Iron1}i), with rapid decrease in the strength of the perturbation as the energy of neutron reaches the 10 eV. Typically, the perturbation in the domains 10 eV-100 eV (Fig. \ref{Fig:Iron1}j), 100eV-1 keV (Fig. \ref{Fig:Iron1}k), and 1-10 keV ((Fig. \ref{Fig:Iron1}l)) are remarked at larger absorber volumes only (r $>$ 5 cm). At energies greater than 2 MeV, i.e. the energy domains 2-3 MeV (Fig. \ref{Fig:Iron2}a), 3-4 MeV (Fig. \ref{Fig:Iron2}b), 4-5 MeV (Fig. \ref{Fig:Iron2}c), 5-6 MeV (Fig. \ref{Fig:Iron2}d), 6-7 MeV (Fig. \ref{Fig:Iron2}e), 7-8 MeV (Fig. \ref{Fig:Iron2}f), 8-9 MeV (Fig. \ref{Fig:Iron2}g), and 9-10 MeV (Fig. \ref{Fig:Iron2}h), the reduction in neutron flux which becomes more pronounced and increases further as the size of the iron absorber increases due to inelastic scattering of iron isotopes.

\begin{figure}[htbp]
  \centering
  \includegraphics[width=0.99\linewidth]{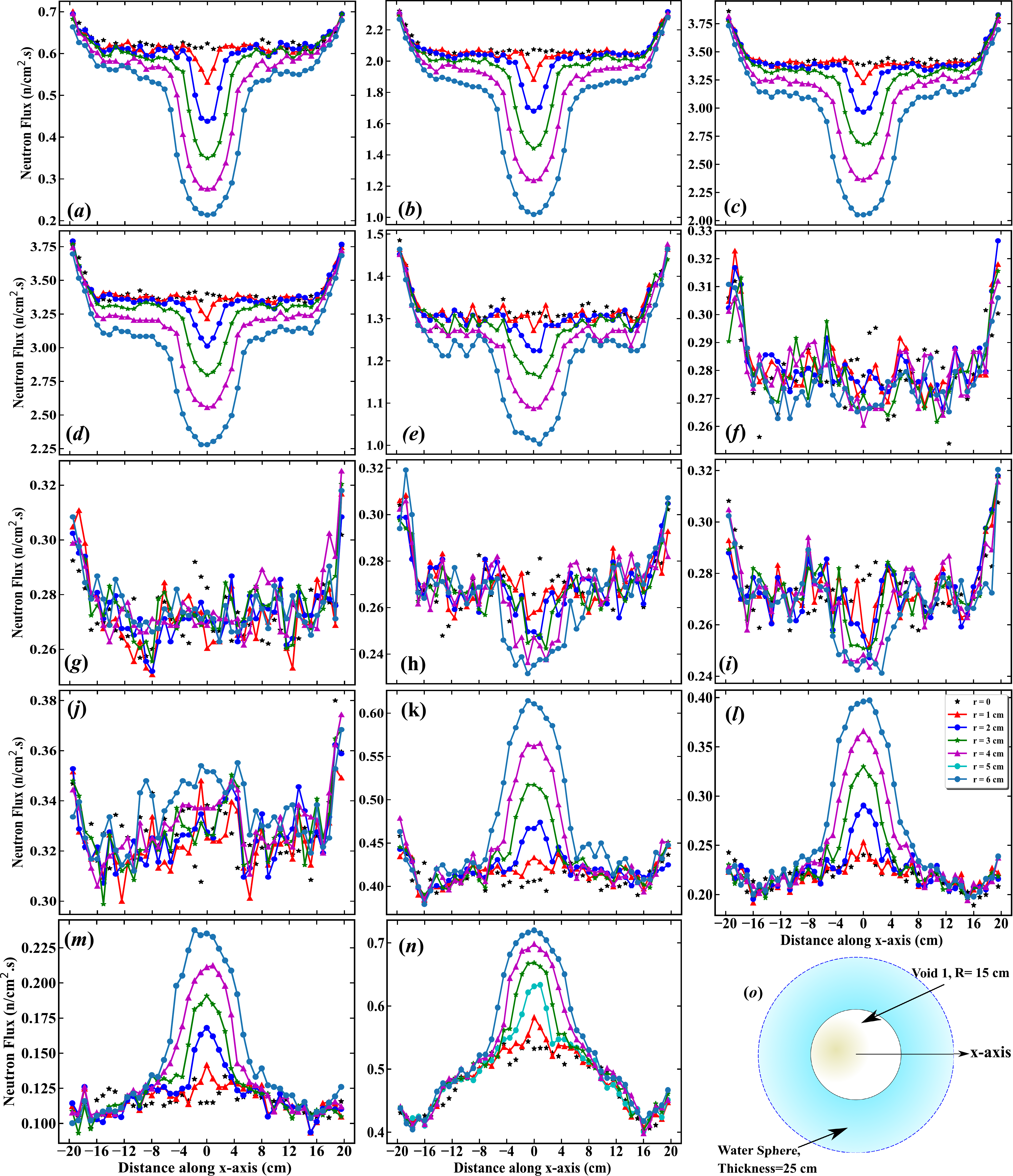}
  \caption{Neutron flux of in the vicinity of zinc absorbers having different dimensions at 14 different domains of energies. Constructed as in Fig. \ref{Fig:Indium1}.}\label{Fig:Zinc1}
\end{figure}

\begin{figure}[htbp]
  \centering
  \includegraphics[width=0.99\linewidth]{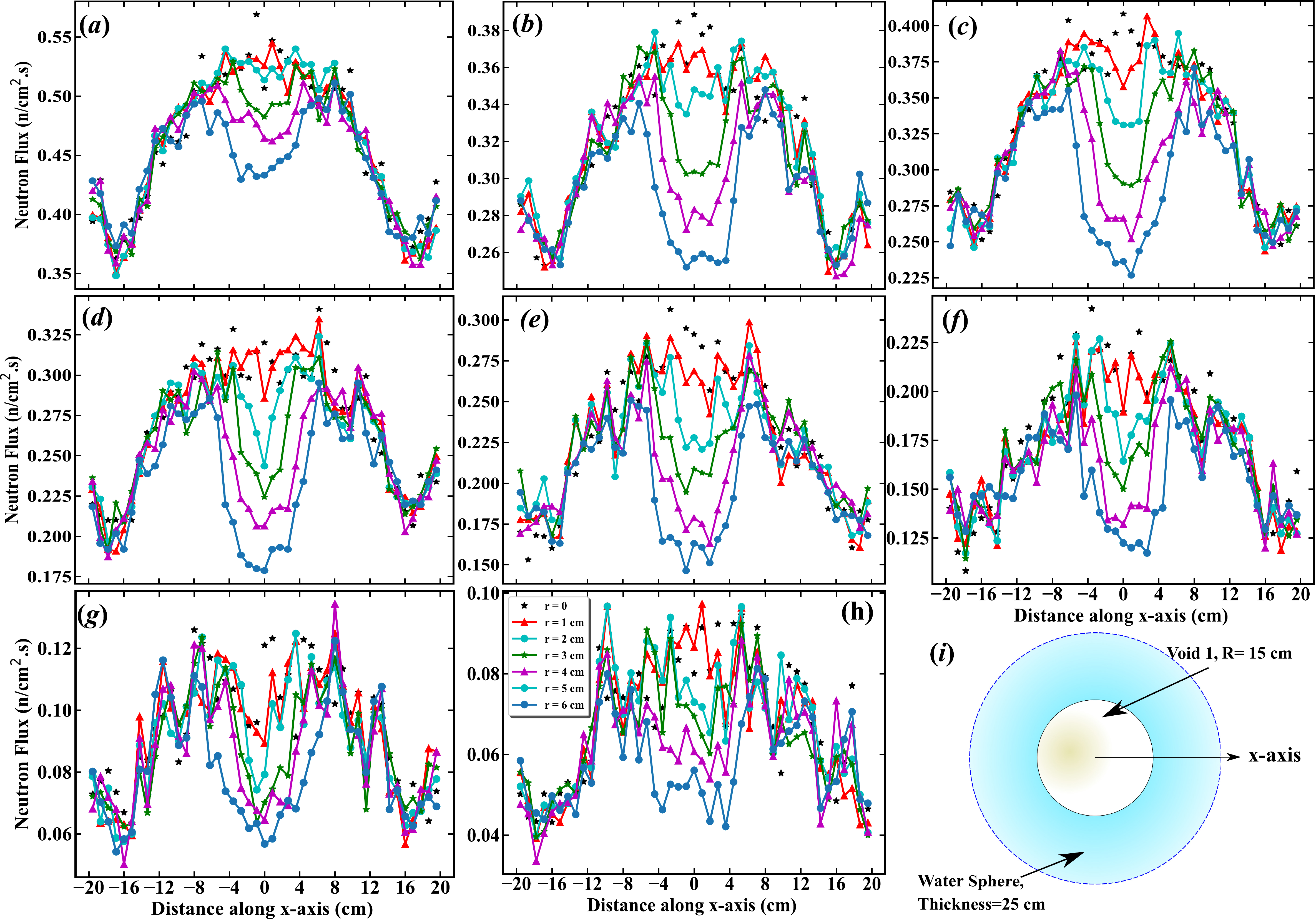}
  \caption{Neutron flux of in the vicinity of zinc absorbers having different dimensions at 8 different domains of energies. Constructed as in Fig. \ref{Fig:Indium2}.}\label{Fig:Zinc2}
\end{figure}

The other weak absorber is zinc isotopes. Zinc has five stable isotopes, i.e. $^{64} $Zn (IA=0.491704, $\sigma _{o} =$0.78 b, $I_{\gamma } =$1.40 b), $^{66} $Zn (IA=0.277306, $\sigma _{o} =$0.61 b, $I_{\gamma } =$0.94b), $^{67} $Zn (IA=0.040 401, $\sigma _{o} =$7.46 b, $I_{\gamma } =$24.1 b), $^{68} $Zn (IA=0.184483, $\sigma _{o} =$1.06 b, $I_{\gamma } =$3.06 b), and $^{68} $Zn (IA=0.006106, $\sigma _{o} =$0.09 b, $I_{\gamma } =$0.10 b), data are from Refs.  \cite{Meija2016293IA,Sukhoruchkin2009book,Mughabghab2006Book,Elmaghraby2019PhysScrCode,Elmaghraby2016Shape}. Zinc, similar to iron, is a weak absorber with strong resonance absorptions between 100 eV and 1 keV. This information is reflected back to the neutron field perturbation induced by zinc as shown in Figs. \ref{Fig:Zinc1} and \ref{Fig:Zinc2}.

Among all of these observations, we noted that the smaller the size of the absorber or with macroscopic cross section, the smaller its induced perturbation, or even negligible.  For a small sample, the change in the flux is a small fraction of the total flux. For this case, an average reaction rate is probably adequate to evaluate the cross-section and resonance integral by the conventional activation method. This case is generally referred to as the infinite dilution limit. The reaction parameters, if it were possible to be detected experimentally, are referred to as the infinite dilution cross-section and resonance integral.

The complexity of the neutron transport phenomenon throws its shadows on every physical system wherever neutron is produced or used. Even a very tiny change in the geometry, which we could think is a small perturbation, affects the neutron energy distribution. The chaotic complexity of neutron transport makes a tiny perturbation in the neutron spectral distribution a severe source of deviation in reaction parameters \cite{Perko201354}. In practical situations,  the most met difficulty in neutron transport is the calculation of spatial and energetic distribution of neutron on stationary conditions in a finite absorbing medium. The general transport equation is a balance equality that conserves number of particles (neutrons) including the possible sources, tracks and losses. Losses includes of accumulation rate of neutrons due to difference in their velocity ($\frac{1}{v} \frac{\partial }{\partial t} n(r,E,\hat{\Omega },t)$), leakage rate out of any volume element ($\hat{\Omega }\cdot \nabla n(r,E,\hat{\Omega },t)$), the total interaction rate including removal due to absorption or scatters out of volume element or change of energy ($\Sigma _{t} (r,E,t)n(r,E,\hat{\Omega },t)$),Where $n(r,E,\hat{\Omega },t)$ is the number density or the density vector of neutron field at specific position, direction and time of neutrons having energy E.  Here, $\hat{\Omega }=\vec{v}/v$ is a unit solid angle vector vector in direction of neutron motion, $\Sigma _{t} (r,E,\hat{\Omega },t)$ is the macroscopic total cross section. Source terms are the rate of neutron production in the volume element which may include fission rate ($\chi _{p} \left(E\right)\int _{0}^{\infty }d E^{{'} } \int _{0}^{\infty }d \Omega ^{{'} } \nu _{p} (E^{{'} } )\Sigma _{f} (r,E^{{'} } ,\Omega ^{{'} } )n(r,E^{{'} } ,\Omega ^{{'} } ,t)$) and delayed neutron emission of the fission products ($\sum _{i=1}^{N}$ $\lambda _{i} \chi _{di}$ $\left(E\right)C_{i} \left(r,t\right) $), where $\chi _{p} \left(E\right)$ and $\chi _{di} \left(E\right)$ refers to the energy spectrum produced in the fission process and delayed neutron decay, respectively; $\nu _{p} $ is the number of neutrons produced per fission; $N$ is the number of delayed neutron sources that could exists at time $t$, $\lambda _{i} $ is the decay rate for each source $i$, $C_{i} \left(r,t\right)$ is the total number of delayed sources $i$ in position $r$  at time $t$.

The main source of neutron density $n(r,E,\hat{\Omega },t)$ at the specific position, energy, direction and time, however exists in the medium, is differential scattering of neutrons from initial energy, $E^{'} $, to the energy, $E$, and from initial direction $\hat{\Omega }^{'} $, to the direction, $\hat{\Omega }$. Which can be expressed as operator $L$ acting on $n$

\begin{widetext}
\begin{equation}\label{Eq.Ln}
\quad Ln=\int_0^{4\pi }d \Omega ^{{'} }  \, \int _{0}^{\infty }d E^{{'} } \Sigma _{s} (r,E^{{'} } \to E,\hat{\Omega }^{{'} } \to \hat{\Omega },t)n(r,E^{{'} } ,\hat{\Omega }^{{'} } ,t)
\end{equation}
\end{widetext}

For infinite source medium having weak absorber within its volume, the balance equality between sources and losses goes stationary (accumulation and leakage rate vanishes) so that
\begin{equation}\label{Eq:Infiniteweak}
\Sigma _{t} (E)n(E)=s(E)+\int _{0}^{\infty }d E^{{'} } \Sigma _{s} (E^{{'} } \to E)n(E^{{'} } )
\end{equation}

The solutions of this equation had been done with several approaches. Horowitz-Tartikoff technique \cite{cadilhac1964neutron} suggested that the spectrum of thermal neutrons in a homogeneous system It applied to heavy moderator and for weak absorbers ($\Sigma _{a} /\zeta \Sigma _{s} $$<$0.1).
\begin{equation}\label{Eq:Horowitz-Tartikoff}
n(E)=C_{N} (M(E)+F(E)).
\end{equation}

Here $M(E)=\frac{E}{\left(kT\right)^{2} } e^{-\; \frac{E}{kT} }$ is the Maxwellian spectrum corresponding to ambient temperature $T$; and F(E) is a perturbation function depends on the thermalization model adopted,  and  $C_{N} $ is a normalization constant adopted to fulfill the condition for unity of total probability. The value of $F(E)$ was analytically derived for first approximation and simple 1/v-law cross section at thermal energies  as \cite{beckurtsWirtz1964}

\begin{widetext}
\begin{equation}\label{Eq:F(E)}
F(E)\approx \frac{\Sigma _{a} (kT)}{\zeta \Sigma _{s} } M(E)\left(\frac{2}{\sqrt{\pi } } \int _{0}^{E/kT}\frac{e^{y} }{y^{2} }  \int _{0}^{y}\sqrt{x}  e^{-x} dxdy-3.19\right)
\end{equation}
\end{widetext}

This is usually called Horowitz-Tartikoff function which has negative contribution to the number density at energies less than 3.2kT and positive contribution at the epithermal energies. For finite pile and high absorbing medium or even inhomogeneous medium, the situation is more complicated and the Horowitz-Tartikoff technique does not applied correctly. Westcott had used the basic idea of foil perturbation applied on the thermal flux was provided by Beckurts  and Wirtz \cite{beckurtsWirtz1964} where the idea that the considerable transport and moderation via scattering are affected by the absorption of neutron by the absorber.  Then, the actual number density of thermal neutrons is less than that would exists in unperturbed filed because of the replacement of none absorbing material by the absorber.  The flux is determined by the average flux on the absorber surface.

Note that in many cases of monoenergetic neutron beam, the absorber induces a small perturbation on the total flux; this case is generally referred to as the infinite dilution limit. However, in field geometry, this causes shift in the flux distribution. In other words, neutron field perturbation is mainly spectral perturbation.

\section{Conclusion}

Using the Monte Carlo N-Particle Transport Code (MCNP), we were able to create a simplified Monte Carlo simulation specifically for a spherical symmetry wrapped by an external neutron source that encloses a moderating volume made of liquid water with a density of 1 g/cm${}^{3}$. Making the source with an {${}^{241}$}AmBe was essential to prove the concept that even when the physics of neutron production is independent of the history of the neutron, there is also a perturbation in the field if an absorber exists. The kernel volume is a void sphere that was proven to contain the uniform neutron distribution up to 1 MeV and has the tendency to be slightly higher at the center for energies greater than 1 MeV due to the spherical symmetry of the geometric pile.  The obtained results showed that any absorber having a density or size greater than the infinite dilution limit would cause neutron field perturbation. The neutron energy distribution is perturbed by even minor changes in the geometry and the type of the material used, either decreasing or increasing depending on the energy domain to satisfy the continuity equation (law of conservation of neutron number). Absorbers of various dimensions perturb the neutron field in a way that is dependent on the absorption and scattering cross-sections, particularly in the neutron resonance area. Due to the chaotic complexity of neutron transport, even tiny perturbation in the spectrum distribution of neutrons can cause a significant deviation in reaction parameters. These include every slight change in the geometry or material used, which we can think a small perturbation, the neutron energy distribution is perturbed. Within the neutron field geometry, the continuity equation required that an absorber within the field should reduce the total number of neutrons in the system.

\section*{Conflicts of interest}
There are no known source for conflicts to declare

%\bibliography{Exciton}

\begin{thebibliography}{26}%
\makeatletter
\providecommand \@ifxundefined [1]{%
 \@ifx{#1\undefined}
}%
\providecommand \@ifnum [1]{%
 \ifnum #1\expandafter \@firstoftwo
 \else \expandafter \@secondoftwo
 \fi
}%
\providecommand \@ifx [1]{%
 \ifx #1\expandafter \@firstoftwo
 \else \expandafter \@secondoftwo
 \fi
}%
\providecommand \natexlab [1]{#1}%
\providecommand \enquote  [1]{``#1''}%
\providecommand \bibnamefont  [1]{#1}%
\providecommand \bibfnamefont [1]{#1}%
\providecommand \citenamefont [1]{#1}%
\providecommand \href@noop [0]{\@secondoftwo}%
\providecommand \href [0]{\begingroup \@sanitize@url \@href}%
\providecommand \@href[1]{\@@startlink{#1}\@@href}%
\providecommand \@@href[1]{\endgroup#1\@@endlink}%
\providecommand \@sanitize@url [0]{\catcode `\\12\catcode `\$12\catcode
  `\&12\catcode `\#12\catcode `\^12\catcode `\_12\catcode `\%12\relax}%
\providecommand \@@startlink[1]{}%
\providecommand \@@endlink[0]{}%
\providecommand \url  [0]{\begingroup\@sanitize@url \@url }%
\providecommand \@url [1]{\endgroup\@href {#1}{\urlprefix }}%
\providecommand \urlprefix  [0]{URL }%
\providecommand \Eprint [0]{\href }%
\providecommand \doibase [0]{https://doi.org/}%
\providecommand \selectlanguage [0]{\@gobble}%
\providecommand \bibinfo  [0]{\@secondoftwo}%
\providecommand \bibfield  [0]{\@secondoftwo}%
\providecommand \translation [1]{[#1]}%
\providecommand \BibitemOpen [0]{}%
\providecommand \bibitemStop [0]{}%
\providecommand \bibitemNoStop [0]{.\EOS\space}%
\providecommand \EOS [0]{\spacefactor3000\relax}%
\providecommand \BibitemShut  [1]{\csname bibitem#1\endcsname}%
\let\auto@bib@innerbib\@empty
%</preamble>
\bibitem [{\citenamefont {Tohamy}\ \emph {et~al.}(2019)\citenamefont {Tohamy},
  \citenamefont {Elmaghraby},\ and\ \citenamefont
  {Comsan}}]{TohamyElmaghrabyComsan2019162387}%
  \BibitemOpen
  \bibfield  {author} {\bibinfo {author} {\bibfnamefont {M.}~\bibnamefont
  {Tohamy}}, \bibinfo {author} {\bibfnamefont {E.~K.}\ \bibnamefont
  {Elmaghraby}},\ and\ \bibinfo {author} {\bibfnamefont {M.}~\bibnamefont
  {Comsan}},\ }\href {https://doi.org/10.1016/j.nima.2019.162387} {\bibfield
  {journal} {\bibinfo  {journal} {Nucl. Instrum. Meth. Phys. Res. A}\ }\textbf
  {\bibinfo {volume} {942}},\ \bibinfo {pages} {162387} (\bibinfo {year}
  {2019})}\BibitemShut {NoStop}%
\bibitem [{\citenamefont {Elmaghraby}\ \emph
  {et~al.}(2019{\natexlab{a}})\citenamefont {Elmaghraby}, \citenamefont
  {Mohamed},\ and\ \citenamefont {Al-abyad}}]{Elmaghraby2019NPA}%
  \BibitemOpen
  \bibfield  {author} {\bibinfo {author} {\bibfnamefont {E.~K.}\ \bibnamefont
  {Elmaghraby}}, \bibinfo {author} {\bibfnamefont {G.~Y.}\ \bibnamefont
  {Mohamed}},\ and\ \bibinfo {author} {\bibfnamefont {M.}~\bibnamefont
  {Al-abyad}},\ }\href {https://doi.org/10.1016/j.nuclphysa.2019.01.009}
  {\bibfield  {journal} {\bibinfo  {journal} {Nucl. Phys.}\ }\textbf {\bibinfo
  {volume} {A984}},\ \bibinfo {pages} {112} (\bibinfo {year}
  {2019}{\natexlab{a}})}\BibitemShut {NoStop}%
\bibitem [{\citenamefont {Al-Abyad}\ \emph {et~al.}(2020)\citenamefont
  {Al-Abyad}, \citenamefont {{El Aal}}, \citenamefont {Hassanin},\ and\
  \citenamefont {Elmaghraby}}]{AlabyadElaalHassaninElmaghraby2020108947}%
  \BibitemOpen
  \bibfield  {author} {\bibinfo {author} {\bibfnamefont {M.}~\bibnamefont
  {Al-Abyad}}, \bibinfo {author} {\bibfnamefont {S.~A.}\ \bibnamefont {{El
  Aal}}}, \bibinfo {author} {\bibfnamefont {W.~F.}\ \bibnamefont {Hassanin}},\
  and\ \bibinfo {author} {\bibfnamefont {E.~K.}\ \bibnamefont {Elmaghraby}},\
  }\href {https://doi.org/10.1016/j.apradiso.2019.108947} {\bibfield  {journal}
  {\bibinfo  {journal} {Appl. Radiat. Isotopes}\ }\textbf {\bibinfo {volume}
  {155}},\ \bibinfo {pages} {108947} (\bibinfo {year} {2020})}\BibitemShut
  {NoStop}%
\bibitem [{\citenamefont {Beckurts}\ and\ \citenamefont
  {Wirtz}(1964)}]{beckurtsWirtz1964}%
  \BibitemOpen
  \bibfield  {author} {\bibinfo {author} {\bibfnamefont {K.~H.}\ \bibnamefont
  {Beckurts}}\ and\ \bibinfo {author} {\bibfnamefont {K.}~\bibnamefont
  {Wirtz}},\ }\href {https://doi.org/10.1007/978-3-642-87614-1} {\emph
  {\bibinfo {title} {Neutron Physics Translated by L. Dresner}}}\ (\bibinfo
  {publisher} {Springer Verlag OHG Berlin},\ \bibinfo {year}
  {1964})\BibitemShut {NoStop}%
\bibitem [{\citenamefont {Walker}\ \emph {et~al.}(1963)\citenamefont {Walker},
  \citenamefont {Randall},\ and\ \citenamefont
  {Stinson~Jr}}]{walker1963thermal}%
  \BibitemOpen
  \bibfield  {author} {\bibinfo {author} {\bibfnamefont {J.~V.}\ \bibnamefont
  {Walker}}, \bibinfo {author} {\bibfnamefont {J.~D.}\ \bibnamefont
  {Randall}},\ and\ \bibinfo {author} {\bibfnamefont {R.~C.}\ \bibnamefont
  {Stinson~Jr}},\ }\href {https://doi.org/10.13182/NSE63-A26442} {\bibfield
  {journal} {\bibinfo  {journal} {Nuclear Science and Engineering}\ }\textbf
  {\bibinfo {volume} {15}},\ \bibinfo {pages} {309} (\bibinfo {year}
  {1963})}\BibitemShut {NoStop}%
\bibitem [{\citenamefont {Elmaghraby}\ \emph
  {et~al.}(2019{\natexlab{b}})\citenamefont {Elmaghraby}, \citenamefont
  {Salem}, \citenamefont {Yousef},\ and\ \citenamefont
  {El-Anwar}}]{Elmaghraby2018PhysScr}%
  \BibitemOpen
  \bibfield  {author} {\bibinfo {author} {\bibfnamefont {E.~K.}\ \bibnamefont
  {Elmaghraby}}, \bibinfo {author} {\bibfnamefont {E.}~\bibnamefont {Salem}},
  \bibinfo {author} {\bibfnamefont {Z.}~\bibnamefont {Yousef}},\ and\ \bibinfo
  {author} {\bibfnamefont {N.}~\bibnamefont {El-Anwar}},\ }\href
  {https://doi.org/10.1088/1402-4896/aaecb0} {\bibfield  {journal} {\bibinfo
  {journal} {Phys. Scr.}\ }\textbf {\bibinfo {volume} {94}},\ \bibinfo {pages}
  {015301} (\bibinfo {year} {2019}{\natexlab{b}})}\BibitemShut {NoStop}%
\bibitem [{\citenamefont {Tohamy}\ \emph {et~al.}(2021)\citenamefont {Tohamy},
  \citenamefont {Elmaghraby},\ and\ \citenamefont
  {Comsan}}]{TohamyElmaghrabyComsan2021045304}%
  \BibitemOpen
  \bibfield  {author} {\bibinfo {author} {\bibfnamefont {M.}~\bibnamefont
  {Tohamy}}, \bibinfo {author} {\bibfnamefont {E.~K.}\ \bibnamefont
  {Elmaghraby}},\ and\ \bibinfo {author} {\bibfnamefont {M.~N.~H.}\
  \bibnamefont {Comsan}},\ }\href {https://doi.org/10.1088/1402-4896/abe258}
  {\bibfield  {journal} {\bibinfo  {journal} {Phys. Scr.}\ }\textbf {\bibinfo
  {volume} {96}},\ \bibinfo {pages} {045304} (\bibinfo {year}
  {2021})}\BibitemShut {NoStop}%
\bibitem [{\citenamefont {Analytis}(1982)}]{analytis1982analysis}%
  \BibitemOpen
  \bibfield  {author} {\bibinfo {author} {\bibfnamefont {G.~T.}\ \bibnamefont
  {Analytis}},\ }\href {https://doi.org/10.1016/0306-4549(82)90093-7}
  {\bibfield  {journal} {\bibinfo  {journal} {Annals of Nuclear Energy}\
  }\textbf {\bibinfo {volume} {9}},\ \bibinfo {pages} {417} (\bibinfo {year}
  {1982})}\BibitemShut {NoStop}%
\bibitem [{\citenamefont {Behringer}\ \emph {et~al.}(1979)\citenamefont
  {Behringer}, \citenamefont {Kos{\'a}ly},\ and\ \citenamefont
  {P{\'a}zsit}}]{behringer1979linear}%
  \BibitemOpen
  \bibfield  {author} {\bibinfo {author} {\bibfnamefont {K.}~\bibnamefont
  {Behringer}}, \bibinfo {author} {\bibfnamefont {G.}~\bibnamefont
  {Kos{\'a}ly}},\ and\ \bibinfo {author} {\bibfnamefont {I.}~\bibnamefont
  {P{\'a}zsit}},\ }\href {https://doi.org/10.13182/NSE79-A20387} {\bibfield
  {journal} {\bibinfo  {journal} {Nuclear Science and Engineering}\ }\textbf
  {\bibinfo {volume} {72}},\ \bibinfo {pages} {304} (\bibinfo {year}
  {1979})}\BibitemShut {NoStop}%
\bibitem [{\citenamefont {Seifritz}(1972)}]{seifritz1972analysis}%
  \BibitemOpen
  \bibfield  {author} {\bibinfo {author} {\bibfnamefont {W.}~\bibnamefont
  {Seifritz}},\ }\href@noop {} {\bibfield  {journal} {\bibinfo  {journal}
  {Atomkernenergie}\ }\textbf {\bibinfo {volume} {19}},\ \bibinfo {pages} {271}
  (\bibinfo {year} {1972})}\BibitemShut {NoStop}%
\bibitem [{\citenamefont {Seifritz}\ and\ \citenamefont
  {Cioli}(1973)}]{seifritz1973load}%
  \BibitemOpen
  \bibfield  {author} {\bibinfo {author} {\bibfnamefont {W.}~\bibnamefont
  {Seifritz}}\ and\ \bibinfo {author} {\bibfnamefont {F.}~\bibnamefont
  {Cioli}},\ }\href@noop {} {\bibfield  {journal} {\bibinfo  {journal} {Trans.
  Amer. Nucl. Soc}\ }\textbf {\bibinfo {volume} {17}},\ \bibinfo {pages} {271}
  (\bibinfo {year} {1973})}\BibitemShut {NoStop}%
\bibitem [{\citenamefont {Laggiard}\ \emph {et~al.}(1995)\citenamefont
  {Laggiard}, \citenamefont {Runkel},\ and\ \citenamefont
  {Stegemann}}]{Laggiard1995124}%
  \BibitemOpen
  \bibfield  {author} {\bibinfo {author} {\bibfnamefont {E.}~\bibnamefont
  {Laggiard}}, \bibinfo {author} {\bibfnamefont {J.}~\bibnamefont {Runkel}},\
  and\ \bibinfo {author} {\bibfnamefont {D.}~\bibnamefont {Stegemann}},\ }\href
  {https://doi.org/10.13182/NSE95-A24113} {\bibfield  {journal} {\bibinfo
  {journal} {Nuclear Science and Engineering}\ }\textbf {\bibinfo {volume}
  {120}},\ \bibinfo {pages} {124} (\bibinfo {year} {1995})}\BibitemShut
  {NoStop}%
\bibitem [{\citenamefont {Antonopoulos-Domis}\ and\ \citenamefont
  {Housiadas}(1999)}]{Antonopoulos-Domis1999337}%
  \BibitemOpen
  \bibfield  {author} {\bibinfo {author} {\bibfnamefont {M.}~\bibnamefont
  {Antonopoulos-Domis}}\ and\ \bibinfo {author} {\bibfnamefont
  {C.}~\bibnamefont {Housiadas}},\ }\href
  {https://doi.org/10.13182/NSE99-A2068} {\bibfield  {journal} {\bibinfo
  {journal} {Nuclear Science and Engineering}\ }\textbf {\bibinfo {volume}
  {132}},\ \bibinfo {pages} {337} (\bibinfo {year} {1999})}\BibitemShut
  {NoStop}%
\bibitem [{\citenamefont {Yamamoto}\ and\ \citenamefont
  {Sakamoto}(2021)}]{Yamamoto2021190}%
  \BibitemOpen
  \bibfield  {author} {\bibinfo {author} {\bibfnamefont {T.}~\bibnamefont
  {Yamamoto}}\ and\ \bibinfo {author} {\bibfnamefont {H.}~\bibnamefont
  {Sakamoto}},\ }\href {https://doi.org/10.1080/00223131.2020.1814176}
  {\bibfield  {journal} {\bibinfo  {journal} {Journal of Nuclear Science and
  Technology}\ }\textbf {\bibinfo {volume} {58}},\ \bibinfo {pages} {190}
  (\bibinfo {year} {2021})}\BibitemShut {NoStop}%
\bibitem [{\citenamefont {Buczk{\'o}}\ and\ \citenamefont
  {Borb{\'e}ly}(1978)}]{buczko1978simple}%
  \BibitemOpen
  \bibfield  {author} {\bibinfo {author} {\bibfnamefont {C.}~\bibnamefont
  {Buczk{\'o}}}\ and\ \bibinfo {author} {\bibfnamefont {A.}~\bibnamefont
  {Borb{\'e}ly}},\ }\href {https://doi.org/10.1007/bf02519415} {\bibfield
  {journal} {\bibinfo  {journal} {Journal of Radioanalytical and Nuclear
  Chemistry}\ }\textbf {\bibinfo {volume} {42}},\ \bibinfo {pages} {393}
  (\bibinfo {year} {1978})}\BibitemShut {NoStop}%
\bibitem [{\citenamefont {Csikai}(1987)}]{csikai1987crc}%
  \BibitemOpen
  \bibfield  {author} {\bibinfo {author} {\bibfnamefont {G.~J.}\ \bibnamefont
  {Csikai}},\ }\href {https://www.osti.gov/biblio/6997209} {\ \textbf {\bibinfo
  {volume} {Volume I}} (\bibinfo {year} {1987})}\BibitemShut {NoStop}%
\bibitem [{\citenamefont {Csikai}\ \emph {et~al.}(2002)\citenamefont {Csikai},
  \citenamefont {Kir{\'a}ly}, \citenamefont {Sanami},\ and\ \citenamefont
  {Michikawa}}]{csikai2002studies}%
  \BibitemOpen
  \bibfield  {author} {\bibinfo {author} {\bibfnamefont {J.}~\bibnamefont
  {Csikai}}, \bibinfo {author} {\bibfnamefont {B.}~\bibnamefont {Kir{\'a}ly}},
  \bibinfo {author} {\bibfnamefont {T.}~\bibnamefont {Sanami}},\ and\ \bibinfo
  {author} {\bibfnamefont {T.}~\bibnamefont {Michikawa}},\ }\href
  {https://doi.org/10.1016/S0168-9002(02)00568-5} {\bibfield  {journal}
  {\bibinfo  {journal} {Nuclear Instruments and Methods in Physics Research
  Section A: Accelerators, Spectrometers, Detectors and Associated Equipment}\
  }\textbf {\bibinfo {volume} {488}},\ \bibinfo {pages} {634} (\bibinfo {year}
  {2002})}\BibitemShut {NoStop}%
\bibitem [{\citenamefont {Thompson}(1979)}]{Thompson1979Mcnp}%
  \BibitemOpen
  \bibfield  {author} {\bibinfo {author} {\bibfnamefont {W.~L.}\ \bibnamefont
  {Thompson}},\ }\href {https://doi.org/10.2172/5519826} {\emph {\bibinfo
  {title} {{MCNP}, a general Monte Carlo code for neutron and photon transport:
  A summary}}},\ \bibinfo {type} {Tech. Rep.}\ \bibinfo {number} {Rep.
  LA---7396-M}\ (\bibinfo  {institution} {Los Alamos National Laboratories,
  USA},\ \bibinfo {year} {1979})\BibitemShut {NoStop}%
\bibitem [{\citenamefont {Tohamy}\ \emph {et~al.}(2020)\citenamefont {Tohamy},
  \citenamefont {Elmaghraby},\ and\ \citenamefont
  {Comsan}}]{TohamyElmaghrabyComsan2020109340}%
  \BibitemOpen
  \bibfield  {author} {\bibinfo {author} {\bibfnamefont {M.}~\bibnamefont
  {Tohamy}}, \bibinfo {author} {\bibfnamefont {E.~K.}\ \bibnamefont
  {Elmaghraby}},\ and\ \bibinfo {author} {\bibfnamefont {M.}~\bibnamefont
  {Comsan}},\ }\href {https://doi.org/10.1016/j.apradiso.2020.109340}
  {\bibfield  {journal} {\bibinfo  {journal} {Appl. Radiat. Isotopes}\ }\textbf
  {\bibinfo {volume} {165}},\ \bibinfo {pages} {109340} (\bibinfo {year}
  {2020})}\BibitemShut {NoStop}%
\bibitem [{\citenamefont {Elmaghraby}(2019)}]{Elmaghraby2019PhysScrCode}%
  \BibitemOpen
  \bibfield  {author} {\bibinfo {author} {\bibfnamefont {E.~K.}\ \bibnamefont
  {Elmaghraby}},\ }\href {https://doi.org/10.1088/1402-4896/ab0845} {\bibfield
  {journal} {\bibinfo  {journal} {Phys. Scr.}\ }\textbf {\bibinfo {volume}
  {94}},\ \bibinfo {pages} {065301} (\bibinfo {year} {2019})}\BibitemShut
  {NoStop}%
\bibitem [{\citenamefont {Elmaghraby}(2017)}]{Elmaghraby2016Shape}%
  \BibitemOpen
  \bibfield  {author} {\bibinfo {author} {\bibfnamefont {E.~K.}\ \bibnamefont
  {Elmaghraby}},\ }\href {https://doi.org/10.1140/epjp/i2017-11516-7}
  {\bibfield  {journal} {\bibinfo  {journal} {Euro. Phys. J. Plus}\ }\textbf
  {\bibinfo {volume} {132}},\ \bibinfo {pages} {249} (\bibinfo {year}
  {2017})}\BibitemShut {NoStop}%
\bibitem [{\citenamefont {Meija}\ \emph {et~al.}(2016)\citenamefont {Meija},
  \citenamefont {Coplen}, \citenamefont {Berglund}, \citenamefont {Brand},
  \citenamefont {Bievre}, \citenamefont {Groning}, \citenamefont {Holden},
  \citenamefont {Irrgeher}, \citenamefont {Loss}, \citenamefont {Walczyk},\
  and\ \citenamefont {Prohaska}}]{Meija2016293IA}%
  \BibitemOpen
  \bibfield  {author} {\bibinfo {author} {\bibfnamefont {J.}~\bibnamefont
  {Meija}}, \bibinfo {author} {\bibfnamefont {T.~B.}\ \bibnamefont {Coplen}},
  \bibinfo {author} {\bibfnamefont {M.}~\bibnamefont {Berglund}}, \bibinfo
  {author} {\bibfnamefont {W.~A.}\ \bibnamefont {Brand}}, \bibinfo {author}
  {\bibfnamefont {P.~D.}\ \bibnamefont {Bievre}}, \bibinfo {author}
  {\bibfnamefont {M.}~\bibnamefont {Groning}}, \bibinfo {author} {\bibfnamefont
  {N.~E.}\ \bibnamefont {Holden}}, \bibinfo {author} {\bibfnamefont
  {J.}~\bibnamefont {Irrgeher}}, \bibinfo {author} {\bibfnamefont {R.~D.}\
  \bibnamefont {Loss}}, \bibinfo {author} {\bibfnamefont {T.}~\bibnamefont
  {Walczyk}},\ and\ \bibinfo {author} {\bibfnamefont {T.}~\bibnamefont
  {Prohaska}},\ }\href {https://doi.org/10.1515/pac-2015-0503} {\bibfield
  {journal} {\bibinfo  {journal} {Pure Appl. Chem.}\ }\textbf {\bibinfo
  {volume} {88}},\ \bibinfo {pages} {293} (\bibinfo {year} {2016})}\BibitemShut
  {NoStop}%
\bibitem [{\citenamefont {Sukhoruchkin}\ \emph {et~al.}(2009)\citenamefont
  {Sukhoruchkin}, \citenamefont {Soroko}, \citenamefont {Gunsing},\ and\
  \citenamefont {Pronyaev}}]{Sukhoruchkin2009book}%
  \BibitemOpen
  \bibfield  {author} {\bibinfo {author} {\bibfnamefont {S.}~\bibnamefont
  {Sukhoruchkin}}, \bibinfo {author} {\bibfnamefont {Z.}~\bibnamefont
  {Soroko}}, \bibinfo {author} {\bibfnamefont {F.}~\bibnamefont {Gunsing}},\
  and\ \bibinfo {author} {\bibfnamefont {V.}~\bibnamefont {Pronyaev}},\
  }\href@noop {} {\emph {\bibinfo {title} {Neutron Resonance Parameters}}},\
  \bibinfo {edition} {1st}\ ed.,\ edited by\ \bibinfo {editor} {\bibfnamefont
  {H.}~\bibnamefont {Schopper}},\ \bibinfo {series} {Landolt-B{\"o}rnstein -
  Group I Elementary Particles, Nuclei and Atoms 24 : Elementary Particles,
  Nuclei and Atoms}, Vol.~\bibinfo {volume} {24}\ (\bibinfo  {publisher}
  {Springer-Verlag Berlin Heidelberg},\ \bibinfo {year} {2009})\BibitemShut
  {NoStop}%
\bibitem [{\citenamefont {Mughabghab}(2006)}]{Mughabghab2006Book}%
  \BibitemOpen
  \bibfield  {author} {\bibinfo {author} {\bibfnamefont {S.~F.}\ \bibnamefont
  {Mughabghab}},\ }\href@noop {} {\emph {\bibinfo {title} {{Atlas of Neutron
  Resonances: Resonance Parameters and Thermal Cross Sections. Z=1-100; 5th
  ed.}}}}\ (\bibinfo  {publisher} {Elsevier},\ \bibinfo {address} {San Diego,
  CA},\ \bibinfo {year} {2006})\BibitemShut {NoStop}%
\bibitem [{\citenamefont {Perk\'{o}}\ \emph {et~al.}(2014)\citenamefont
  {Perk\'{o}}, \citenamefont {Gilli}, \citenamefont {Lathouwers},\ and\
  \citenamefont {Kloosterman}}]{Perko201354}%
  \BibitemOpen
  \bibfield  {author} {\bibinfo {author} {\bibfnamefont {Z.}~\bibnamefont
  {Perk\'{o}}}, \bibinfo {author} {\bibfnamefont {L.}~\bibnamefont {Gilli}},
  \bibinfo {author} {\bibfnamefont {D.}~\bibnamefont {Lathouwers}},\ and\
  \bibinfo {author} {\bibfnamefont {J.~L.}\ \bibnamefont {Kloosterman}},\
  }\href {https://doi.org/10.1016/j.jcp.2013.12.025} {\bibfield  {journal}
  {\bibinfo  {journal} {J. Comput. Phys.}\ }\textbf {\bibinfo {volume} {260}},\
  \bibinfo {pages} {54} (\bibinfo {year} {2014})}\BibitemShut {NoStop}%
\bibitem [{\citenamefont {Cadilhac}\ \emph {et~al.}(1964)\citenamefont
  {Cadilhac}, \citenamefont {Soule},\ and\ \citenamefont
  {Tretiakoff}}]{cadilhac1964neutron}%
  \BibitemOpen
  \bibfield  {author} {\bibinfo {author} {\bibfnamefont {M.}~\bibnamefont
  {Cadilhac}}, \bibinfo {author} {\bibfnamefont {J.-L.}\ \bibnamefont
  {Soule}},\ and\ \bibinfo {author} {\bibfnamefont {O.}~\bibnamefont
  {Tretiakoff}},\ }\href@noop {} {\emph {\bibinfo {title} {Neutron
  thermalization and spectra}}},\ \bibinfo {type} {Tech. Rep.}\ (\bibinfo
  {institution} {CEA Saclay},\ \bibinfo {year} {1964})\BibitemShut {NoStop}%
\end{thebibliography}

\end{document}